\documentclass[aps,pra,twocolumn,floatfix,10pt]{revtex4-1}
\usepackage[utf8]{inputenc}
\usepackage{framed}
\usepackage{graphicx}
\usepackage{amsmath}
\usepackage{amsfonts}
\usepackage{framed}
\usepackage{subfigure}
\usepackage{hyperref}

\hypersetup{
  colorlinks   = true, 
  urlcolor     = blue, 
  linkcolor    = blue, 
  citecolor   = red, 
}

\newcommand{\bra}[1]{\mbox{$\langle #1 |$}}
\newcommand{\ket}[1]{\mbox{$| #1 \rangle$}}
\newcommand{\braket}[2]{\mbox{$\langle #1 | #2 \rangle$}}

\newcommand{\tr}{\mbox{$\text{tr}$}}

\def\dee{{\mathrm{d}}}
\def\ic{{\mathrm{i}}}

\def\CFT{\mbox{\tiny CFT}}

\def\aUV{a_{\mbox{\tiny UV}}}
\def\inn{\mbox{\tiny in}}
\def\out{\mbox{\tiny out}}
\def\H2{\mbox{\tiny H$_2$}}
\def\E2{\mbox{\tiny E$_2$}}
\def\TN{\mbox{\tiny TN}}

\allowdisplaybreaks[3]
\def\letter{paper } 
\def\appendix{Appendix }

\begin{document}

\title{Tensor networks as path integral geometry}
\author{Ashley Milsted}
\affiliation{Perimeter Institute for Theoretical Physics, Waterloo, Ontario N2L 2Y5, Canada}  
\author{Guifre Vidal}
\affiliation{Perimeter Institute for Theoretical Physics, Waterloo, Ontario N2L 2Y5, Canada}  \date{\today}

\begin{abstract}
In the context of a quantum critical spin chain whose low energy physics corresponds to a conformal field theory (CFT), it was recently demonstrated [A. Milsted G. Vidal, arXiv:1805.12524] that certain classes of tensor networks used for numerically describing the ground state of the spin chain can also be used to implement (discrete, approximate versions of) conformal transformations on the lattice. In the continuum, the same conformal transformations can be implemented through a CFT path integral on some curved spacetime. Based on this observation, in this \letter we propose to interpret the tensor networks themselves as a path integrals on curved spacetime. This perspective assigns (a discrete, approximate version of) a geometry to the tensor network, namely that of the underlying curved spacetime. 
\end{abstract}


\maketitle

The tensor network formalism \cite{MPS,MERA,PEPS,stat1,stat2,stat3} allows for an efficient description of complex objects, such as many-body wavefunctions in quantum systems, and has in recent years become useful in a wide range of disciplines, including condensed matter \cite{MPS,MERA,PEPS}, statistical mechanics \cite{stat1,stat2,stat3}, quantum gravity \cite{H2,dS2,EHM,HAPPY,Random}, quantum chemistry \cite{QChem}, quantum and classical information theory \cite{QErrorC}, and machine learning \cite{MachineLearning}.
The geometry of a tensor network is of central importance from a computational viewpoint, since it largely determines our ability to perform efficient calculations.
Moreover, the geometry also dictates the structure of possible correlations or dependencies within the complex object that the tensor network represents, e.g. the asymptotic scaling of entanglement entropy in many-body wavefunctions \cite{geometry}. Finally, geometry is particularly important for the holographic interpretation of tensor networks in the context of the anti de Sitter / conformal field theory (AdS/CFT) correspondence \cite{AdSCFT}. There, tensor networks with hyperbolic geometry, such as the \textit{multi-scale entanglement renormalization ansatz} (MERA) \cite{MERA,H2,dS2}, the \textit{exact holographic mapping} \cite{EHM}, the \textit{holographic error correction code} \cite{HAPPY} or the \textit{random tensor networks} of Ref.\ \cite{Random}, have been proposed as toy models for quantum gravity in AdS background. 
However, it is not always clear how to properly assign a geometry to a tensor network, or what that assignment really means. As a matter of fact, and rather confusingly, MERA has been speculated to realize two mutually incompatible geometries: the hyperbolic plane H$_2$ \cite{H2} and de Sitter spacetime dS$_{2}$ \cite{dS2}. 

In this \letter we propose a geometric characterization of a specific class of tensor networks, used to describe critical quantum spin chains, by arguing that they can be naturally interpreted as a CFT path integral on curved spacetime, see also \cite{PathIntegral1,PathIntegral2,TNRlocal}. Specifically, we consider two-dimensional tensor networks composed of three types of tensors: \textit{euclideons} $e$, \textit{disentanglers} $u$, and \textit{isometries} $w$ (in absence of translation invariance, additional tensors known as \textit{smoothers} are also included). Euclideons $e$ are obtained from an euclidean time evolution by the Hamiltonian $H$ of the critical quantum spin chain (or Boltzmann weights of a related two-dimensional statistical partition function), whereas disentanglers $u$ and isometries $w$ are MERA tensors optimized to represent the ground state of $H$ and can be obtained from the euclideons $e$ \cite{TNRMERA}. We recently demonstrated that strips of such tensor networks define linear maps $V$ that act on the low energy states of the quantum spin chain as conformal maps \cite{Conformal}. Each such conformal maps can be implemented in the continuum by a CFT path integral on a strip of some curved spacetime. By analogy, here we argue that we can then think of such tensor networks as describing (a discrete version of) a CFT euclidean path integral on a two-dimensional curved spacetime. In this way, we can assign a geometry to the tensor network: that of the underlying curved spacetime. However, a CFT path integral is invariant under local (Weyl) scale transformations. Hence requiring compatibility with a CFT path integral only determines the tensor network geometry up to a \textit{local} scale factor $\Omega(\tau,x )$. A unique geometry (up to a \textit{global} scale factor, or choice of short distance cut-off $\aUV$) can then be assigned by adding a second natural requirement, namely that the proper distance between nearest neighbor tensors be constant throughout the network. Accordingly, an euclideon $e$ anywhere in the network is seen to represent the euclidean path integral on a square patch of flat spacetime of size $\aUV \times \aUV$, whereas disentanglers $u$ and isometries $w$ implement space diffeomorphisms (local coordinate rescaling). A discrete curved geometry is then composed of microscopic square patches of flat euclidean spacetime connected non-trivially by such diffeomorphisms. Finally, as an example demonstrates, a systematically improved approximation to a continuous geometry is obtained by decreasing the cut-off $\aUV$. 

\begin{figure}
\includegraphics[width=\linewidth]{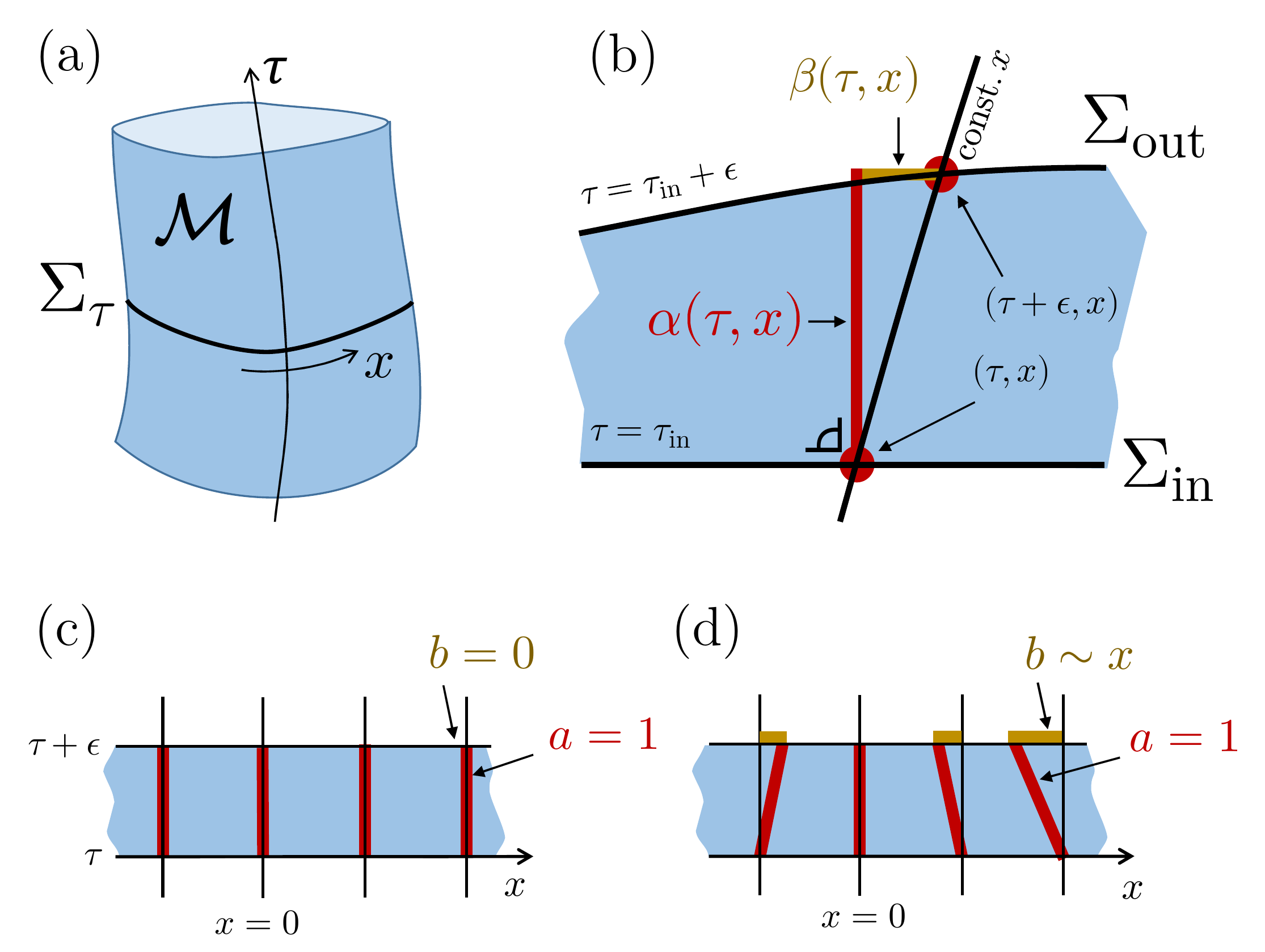}
\caption{
(a) Manifold $\mathcal{M}$ that can be foliated into euclidean time slices $\Sigma_{\tau}$.
(b) The transition between time slices $\Sigma_{\inn}$ and $\Sigma_{\out}$, where $\tau_{\out} - \tau_{\inn} = \epsilon \ll 1$, is characterized by \textit{lapse} and \textit{shift} functions $\alpha(\tau,x)$ and $\beta(\tau,x)$ that are proportional to the profile functions $a(\tau,x)$ and $b(\tau,x)$ in Eqs.\ \eqref{eq:Q0} and \eqref{eq:Q1}, respectively (see appendix.~\ref{sec:AppI}). (c) Profile functions $a=1$ and $b=0$ for the euclidean plane E$_2$ in the coordinates $\tau,x$ of Eq.~\eqref{eq:gE2}. (d) Profile functions $a=1$ and $b=x/L$ for the hyperbolic plane H$_2$ in the coordinates $\tau,x$ of Eq.~\eqref{eq:gH2}.
\label{fig:geometry} 
}
\end{figure}

\textit{Euclidean path integral as a linear map.---} Let $(\mathcal{M},g)$ denote a two-dimensional euclidean spacetime where the spacetime manifold $\mathcal{M}= \bigcup_{\tau} \Sigma_{\tau}$ can be decomposed into `euclidean time' slices $\Sigma_\tau$, each topologically equivalent to a fixed one-dimensional `space' manifold $\Sigma$ (see Fig.~\ref{fig:geometry}(a)). A generic metric $g$ in coordinates $x^{\mu} = (\tau,x)$ reads
\begin{equation} \label{eq:g}
g_{\mu\nu}(\tau,x) = \Omega^{2}(\tau,x)\left(\begin{array}{cc} A(\tau,x)^2 & B(\tau,x)\\ B(\tau,x) & 1 \end{array} \right),
\end{equation}
where $\Omega(\tau,x)$ is a scale factor and $A(\tau,x)$ and $B(\tau,x)$ are general functions such that $A(\tau,x)^2>B(\tau,x)^2$. A CFT is defined on this spacetime as usual in terms of a field (or set of fields) $\phi(\tau,x)$ and its path integral
\begin{equation}
\mathcal{Z} = \int [D\phi]~ e^{-S[\phi]},~~~~~ S \equiv \int_{\mathcal{M}}\!\!\! d\tau\, dx
~\sqrt{|g|} 
~\mathcal{L}\left(\phi(\tau,x)\right),
\end{equation}
where $S$ and $\mathcal{L}$ are the CFT action and lagrangian density. Crucial to our discussion, given a spacetime strip $\mathcal{N}\subset \mathcal{M}$ delimited by slices $\Sigma_{\inn}$ and $\Sigma_{\out}$ for times $\tau_{\inn}$ and $\tau_{\out}$, the path integral restricted to $\mathcal{N}$ defines a linear map $V$ in the CFT Hilbert space $\mathcal{H}(\Sigma)$, see \cite{SupplMat} for a review. Specifically, $V$ has matrix elements
\begin{eqnarray} \label{eq:V1}
\bra{\varphi_{\out}(x)} V \ket{\varphi_{\inn}(x)} = \int [D\phi]~ e^{-S_{\mathcal{N}}[\phi]},
\end{eqnarray}
where the field configurations $\phi(\tau,x)$ are integrated over $\mathcal{N}$ with $\varphi_{\inn}(x)$ and $\varphi_{\out}(x)$ as boundary conditions. When the strip is thin, that is $\tau_{\out} = \tau_{\inn}+\epsilon$ with $\epsilon \ll 1$ (Fig.\ \ref{fig:geometry}(b)), we can expand $V= e^{-\epsilon Q} \approx \mathbb{I} - \epsilon Q$ to linear order in $\epsilon$.  Its generator $Q = Q_0 + iQ_1$ has two contributions 
\begin{eqnarray}
Q_0 &\equiv& \int_{\Sigma_{\inn}} \!\!dx~a(\tau,x) ~h(\tau,x), \label{eq:Q0}\\
Q_1 &\equiv& \int_{\Sigma_{\inn}} \!\!dx~b(\tau,x) ~p(\tau,x), \label{eq:Q1}
\end{eqnarray}
that depend on the CFT hamiltonian and momentum densities $h(\tau,x)$ and $p(\tau,x)$ as well as on the metric $g_{\mu\nu}(\tau,x)$ in Eq.~\eqref{eq:g} through  
\begin{equation} \label{eq:abAB}
a(\tau,x) = \sqrt{A(\tau,x)^2-B(\tau,x)^2}, ~~~b(\tau,x) = B(\tau,x). 
\end{equation}
Here $Q_0$ is the generator of a \textit{non-uniform time evolution} in $\mathcal{H}(\Sigma)$, corresponding to a geometric deformation normal to the slice $\Sigma_{\inn}$ and proportional to $a(x)$; $Q_1$ generates instead a \textit{non-uniform space translation} or \textit{local rescaling}, corresponding to a reparametrization of space within $\Sigma_{\inn}$ according to $b(x)$, see Fig.\ \ref{fig:geometry}(b). 

\textit{Examples.---} A generator of type $Q_0$ is the Hamiltonian
\begin{equation} \label{eq:H}
H \equiv \int_{\Sigma_{\inn}} dx~h(\tau,x),
\end{equation}
which corresponds to $a(\tau,x)=1$ and generates uniform time evolution in e.g. flat spacetime. Familiar $Q_1$ generators include the momentum operator $P \equiv \int_{\Sigma_{\inn}} dx~p(\tau,x)$, which corresponds to $b(\tau,x)=1$ and generates uniform space translations $x\rightarrow x+\epsilon$, and the dilation operator
\begin{equation} \label{eq:D}
 D \equiv \int_{\Sigma_{\inn }} dx~x~p(\tau,x),
\end{equation} 
which corresponds to $b(\tau,x)=x$ and generates a uniform rescaling $x \rightarrow x +\epsilon x$ of the space coordinate. 
For later reference, we consider two simple geometries explicitly: 

(i) The \textit{euclidean plane} E$_2$ has metric 
\begin{equation} \label{eq:gE2}
g^{\E2}_{\mu\nu}(\tau,x) = \left(\begin{array}{cc} 1 & 0\\ 0 & 1 \end{array} \right),
\end{equation}
implying $\Omega(\tau,x)= 1 = A(\tau,x) = a(\tau,x)$ and $B(\tau,x)=b(\tau,x)=0$ above. Therefore the path integral on a strip of with $\epsilon$ produces a linear map $V=e^{-\epsilon H}$, generated by the Hamiltonian $H$ of Eq.~\eqref{eq:H}, see Fig.\ \ref{fig:geometry}(c). 

(ii) The \textit{hyperbolic plane} H$_2$ with radius of curvature $L$ has metric $\tilde{g}^{\H2}_{\mu\nu}(\eta,r) = \left(L/\eta\right)^2\delta_{\mu\nu}$ in some conformal coordinates $\eta,r$ (with $\eta\geq 0$ and $r\in \mathbb{R}$).
With an alternative choice of coordinates $\tau \equiv L\log(\eta/L)$ and $x \equiv rL/\eta$ (with $\tau,x \in \mathbb{R}$), the same metric reads
\begin{equation} \label{eq:gH2}
g^{\H2}_{\mu\nu}(\tau,x) = \left(\begin{array}{cc} 1+(x/L)^2 & x/L \\ x/L & 1 \end{array} \right),
\end{equation}
or $\Omega(\tau,x)=1$, $A(\tau,x)^2=1+(x/L)^2$, $B(\tau,x) = x/L$ in Eq.~\eqref{eq:g}; $a(\tau,x) = 1$ and $b(\tau,x) = x/L$ in Eqs.~\eqref{eq:Q0}-\eqref{eq:Q1}. Thus in these coordinates the path integral on a thin strip of H$_2$ implements a conformal map $V=e^{-\epsilon Q^{\H2}}$ with generator
\begin{equation} \label{eq:QH2}
Q^{\H2} = \int dx~\left( h(\tau,x) +\frac{ix}{L} p(\tau,x) \right) = H+ \frac{i}{L}D,
\end{equation}
namely a linear combination of the Hamiltonian $H$ and the dilation operator $D$ in Eqs.\ \eqref{eq:H}-\eqref{eq:D}, see Fig.\ \ref{fig:geometry}(d). 

\textit{Reconstructing the metric from the path integral.---} Suppose now that, for all possible values of time $\tau_{\inn}$, we have access to the infinitesimal conformal map $V=e^{-\epsilon Q}$ obtained from the path integral on a thin strip at time $\tau_{\inn}$. Do we have enough information to reconstruct the metric $g_{\mu\nu}(\tau,x)$ in Eq.~\eqref{eq:g}? By inspecting the generator $Q$ and inverting Eq.~\eqref{eq:abAB}, we can certainly recover $A(\tau,x)$ and $B(\tau,x)$. This determines the conformal class of the metric, from which non-trivial information, such as its signature, can already be extracted  \cite{geoMERA}. However the method fails to provide the local scale factor $\Omega(\tau,x)$ in Eq.~\eqref{eq:g}. This was to be expected, since the path integral of a CFT is essentially invariant under local (Weyl) rescaling \cite{Weyl}. In order to completely specify the geometry we must additionally provide the scale factor $\Omega(\tau,x)$. Our proposed assignment of geometry to tensor networks (rule 1 below), will run into the same difficulty, which will be resolved by additionally specifying a scale factor (rule 2 below).  

\begin{figure}
\includegraphics[width=\linewidth]{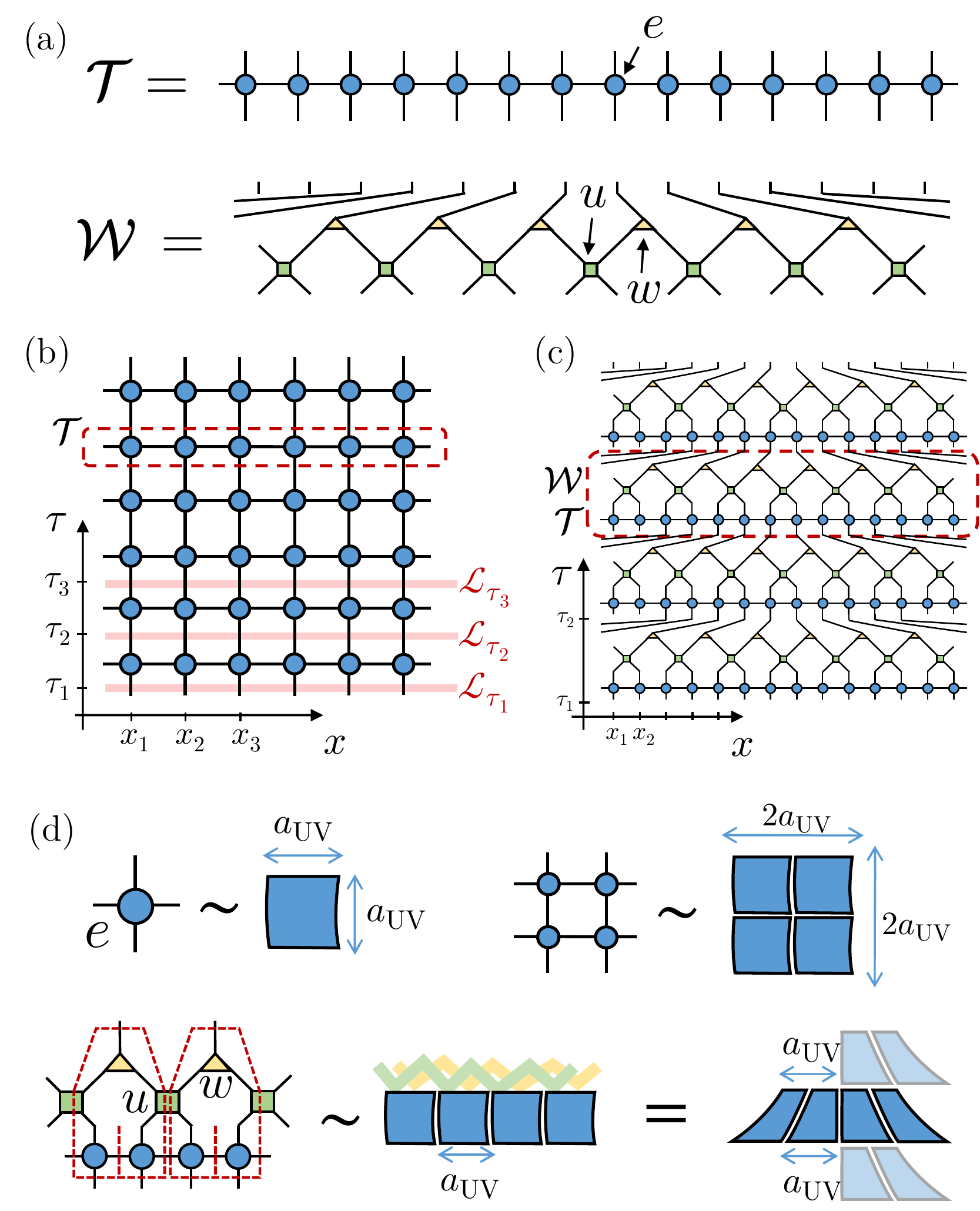}
\caption{
(a) Layer $\mathcal{T}$ of euclideons $e$ that implements a uniform euclidean time evolution $e^{-H}$ and double layer $\mathcal{W}$ of disentanglers $u$ and isometries $w$ that implements a uniform rescaling of space $2^{-iD}$.
(b) Tensor network TN$_1 = \mathcal{T}^{n}$ implementing $e^{-nH}$, which by rules 1-2 corresponds to a path integral on flat euclidean spacetime E$_2$.
(c) Tensor network TN$_2 = (\mathcal{W}\mathcal{T})^n$ implementing $2^{ -n(H+iD)}$, which by rules 1-2 corresponds to a path integral on the hyperbolic plane H$_2$.
(d) An euclideon $e$ represents a (path integral on a) $\aUV \times \aUV$ square patch of euclidean spacetime. By contracting several euclideons together, we obtain a larger patch of flat spacetime. A layer $\mathcal{W}$ of disentanglers and isometries allows us to glue the top of two euclideons with the bottom of one euclideon.
\label{fig:TN}
}
\end{figure}

\textit{Tensor networks as geometry.---} Consider now a critical quantum spin chain whose low energy, long-distance physics is described by a CFT. Ref.\ \cite{Conformal} recently argued that certain tensor networks implemented lattice versions of conformal transformations on the low energy states of the quantum spin chain. The simplest examples, see Figs. \ref{fig:TN}(a), are a single layer $\mathcal{T}$ of \textit{euclideons} $e$, which implements an euclidean time evolution $e^{-H}$, and a double layer $\mathcal{W}$ of disentanglers $u$ and isometries $w$, which implements a global rescaling $2^{-iD}$. More generally, one can combine euclideons, disentanglers, and isometries (and auxiliary tensors called smoothers \cite{Conformal}) into a single strip that is seen to implements a more general, finite linear map $e^{-Q}$, where $Q = Q_0 +i Q_1$ generates both non-uniform time evolution and local rescaling.


In this \letter we propose to interpret this class of tensor networks as a path integral on a curved spacetime. Given a tensor network, we first decompose it as a product of layers or strips. For instance, the tensor network TN$_1$ of Fig.\ \ref{fig:TN}(b) decomposes as the product of strips $\mathcal{T}$, whereas the tensor network TN$_2$ of Fig.\ \ref{fig:TN}(d) decomposes as the product of strips $\mathcal{W}\mathcal{T}$. Each strip has two time slices as its boundaries. Each time slice corresponds to some discrete value $\tau_m$ of euclidean time $\tau$ and is equipped with a lattice $\mathcal{L}_{\tau_m}$ whose sites are labeled by discrete space-time coordinates $(\tau_m,x_n)$ and correspond to the indices cut by that time slice, see e.g. Fig.\ \ref{fig:TN}(b). Lattice $\mathcal{L}_{\tau_m}$ hosts the Hilbert space of a quantum spin chain, with each cut index representing a quantum spin. Then a strip of the tensor network implements a linear map between the spin chains at its bottom and top boundaries. We assign a geometry or metric to a tensor network by assigning a metric to each of its strips, following two rules.

\textit{Rule 1 (compatibility with path integral).---} Let $e^{-Q}$ be the linear map implemented by a strip of a tensor network. Then the assigned strip geometry/metric should be such that the CFT path integral on it also produces the linear map $e^{-Q}$. 


For instance, since a layer $\mathcal{T}$ of TN$_1$ implements the linear map $e^{-H}$, with $a(\tau,x)=1$ and $b(\tau,x)=0$ in Eqs.\ \eqref{eq:Q0}-\eqref{eq:Q1} or $A(\tau,x)^2 = 1$ and $B(\tau,x)=0$ in Eq.\ \eqref{eq:g}, following rule 1 we assign the metric \cite{rule1}
\begin{equation} \label{eq:gTN1}
g_{\mu\nu}^{\TN_1}(\tau,x) = \Omega^2(\tau,x) \left(\begin{array}{cc} 1 & 0\\ 0 & 1 \end{array} \right)
\end{equation}
to tensor network TN$_1$. In turn, each strip of tensor network TN$_2$ implements the linear map $\mathcal{W}\mathcal{T} =2^{-iD}e^{-H} = 2^{- H- iD}$ \cite{SupplMat}, which equals $e^{-sQ}$ for $s=\log(2)$ and generator $Q=H+iD$. This generator is precisely $Q^{\H2}$ in Eq.~\eqref{eq:QH2} for $L=1$, corresponding to $a(\tau,x)=1$ and $b(\tau,x) = x$, or $A(\tau,x)^2 = 1+x^2$ and $B(\tau,x)=x$. Accordingly, rule 1 assigns to TN$_2$ the metric \cite{rule1} 
\begin{equation} \label{eq:gTN2}
g_{\mu\nu}^{\TN_2}(\tau,x) = \Omega^2(\tau,x) \left(\begin{array}{cc} 1+x^2 & x \\ x & 1 \end{array} \right).
\end{equation}
We emphasize that rule 1 does not determine the scale factor $\Omega(\tau,x)$ in $g^{\TN_1}_{\mu\nu}$ or $g^{\TN_2}_{\mu\nu}$, since \textit{any} scale factor $\Omega(\tau,x)$ is compatible with the linear maps $e^{-H}$ and $2^{-H-iD}$ implemented by a strip of these networks \cite{Weyl}. 

\textit{Rule 2 (constant lattice spacing).---} The proper distance between any two equal-time, nearest neighbor lattice sites $(\tau_{m},x_n)$ and $(\tau_{m},x_{n+1})$ is a constant $\aUV$ for all discrete values of $\tau_m$ and $x_n$ across the tensor network. 

For example, when applied to the tensor networks TN$_1$ and TN$_2$ above, rule 2 sets the scale factor of the metrics $g^{\TN_1}$ and $g^{\TN_2}$ in Eqs.\ \eqref{eq:gTN1}-\eqref{eq:gTN2} to $\Omega(\tau,x) = \aUV$. Temporarily choosing $\aUV=1$ for simplicity, we then identify the geometry of these tensors networks as the Euclidean plane E$_2$ in Eq.\ \eqref{eq:gE2} and the hyperbolic plane H$_2$ (with radius $L=1$) in Eq.\ \eqref{eq:gH2}, respectively. 

It follows from the above examples that, as illustrated in Fig.\ \ref{fig:TN}(d), we can think of each euclideon $e$ as representing (a path integral on) a square patch of flat spacetime of size $\aUV \times \aUV$. In contrast, disentanglers and isometries do not represent a patch of geometry, but allow us to glue together the path integral on square pieces of flat spacetime (as given by the euclideons $e$) non-trivially into a path integral on some curved spacetime, such as H$_2$.

\begin{figure}
  \scriptsize
  \raggedright (a)\\ 
  \includegraphics[width=\linewidth]{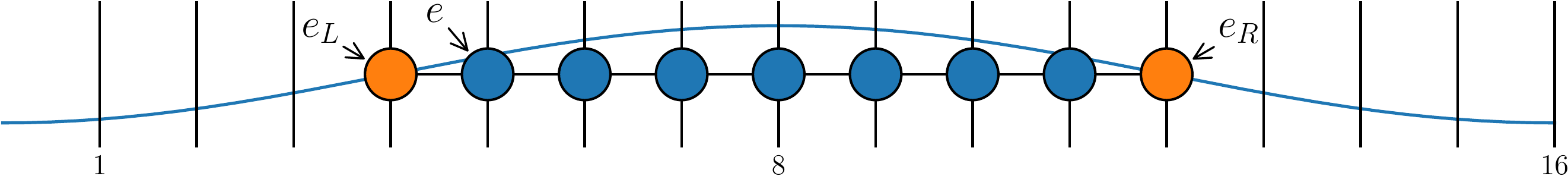}\\
  \includegraphics[width=\linewidth]{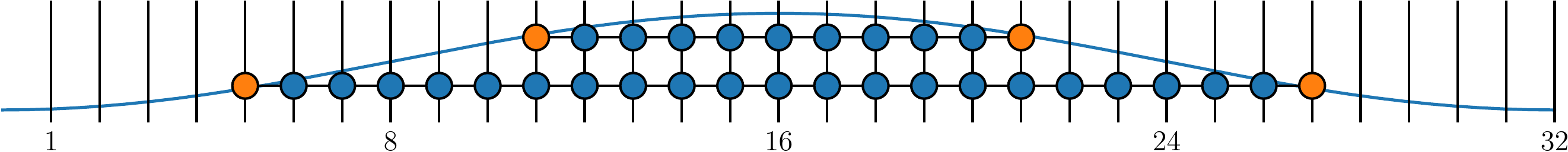}\\
  \includegraphics[width=\linewidth]{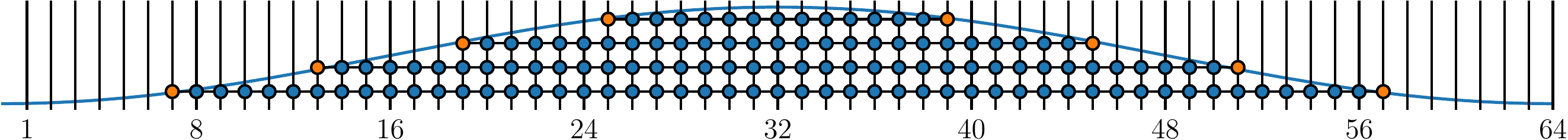}\\
  (b)\\
  \vspace{0.1cm}
  \includegraphics[width=\linewidth]{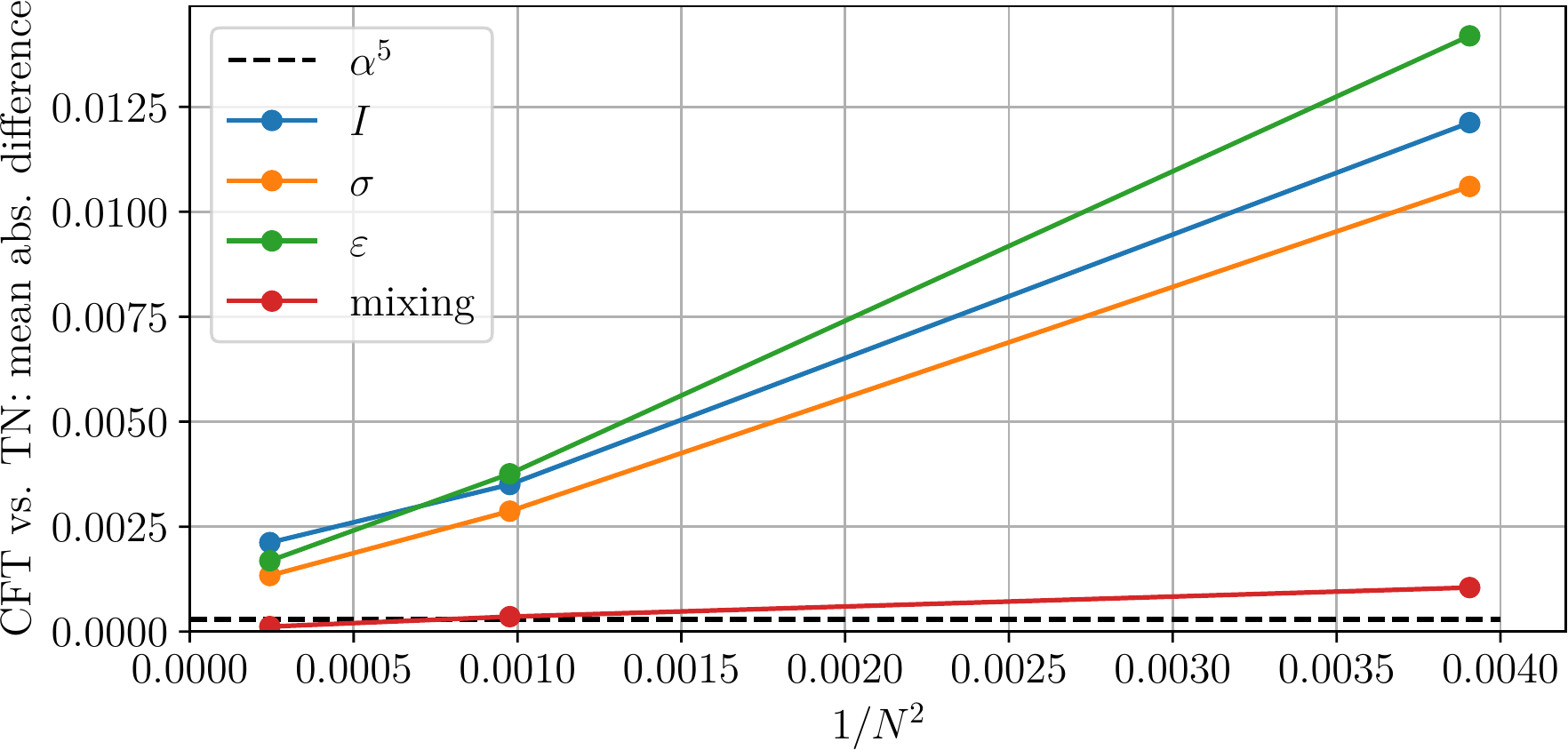}
\caption{
(a) Sequence of tensor networks, made of euclideons $e$ and smoothers $e_L$ and $e_R$, that approximate the continuous profile $f(x) = \alpha(1-\cos\left(x\right))$, with $\alpha=2\pi/32$. The x-axis labels sites $1$ to $N$ with lattice spacing $2\pi/N$. The corresponding linear maps $V_N$ are expected to act as improving approximations to $V=e^{-Q}$. 
(b) The difference between the numerical matrix elements $\bra{\phi_\alpha} V_N \ket{\phi_{\beta}}$, for the first 41 low-energy states of the quantum critical spin chain (critical transverse field Ising model), and the corresponding \cite{CFT} CFT matrix elements of $V$. We plot the mean absolute difference within each conformal tower, as well as between towers (mixing). We observe convergence in $N$ up to an error $\mathcal{O}(\alpha^5)$ made in computing the matrix elements of $V$ to order $\alpha^4$. For these computations we used periodic Matrix Product States \cite{MPS, MPS_crit, pMPS, puMPS} and lattice Virasoro generators \cite{LatVira}. See appendix~\ref{sec:app_refinement}.
\label{fig:refinement} 
}
\end{figure}

\textit{Refinement limit.---} Given a curved spacetime with metric $g_{\mu\nu}$ as a target, we can use a sequence of tensor networks with decreasing lattice spacing $\aUV$ to systematically obtain a finer approximation to the CFT path integral on the target continuous spacetime. As an example, consider a topological cylinder $\mathcal{M} = \mathbb{R}\times S^{1}$, where each time slice $\Sigma_{\tau}$ for $\tau \in \mathbb{R}$ is now a circle $S^{1}$ of unit perimeter and, thus parameterized, the space coordinate $x\in [0,2\pi)$ is periodic, and let the metric $g_{\mu\nu}$ be as in Eq.~\eqref{eq:g} with $\Omega(\tau,x)=1$, $A(\tau,x) = f(x)^2$ and $B(\tau,x)=0$, where $f(x) \equiv \alpha(1-\cos\left(x\right))$. Fig.\ \ref{fig:refinement}(d) shows a sequence of discrete approximations to a periodic strip with boundaries $\tau_{\inn}=0$ and $\tau_{\out}=1$, which has an $x$-dependent width (as measured in proper euclidean time) given by $f(x)$. In the continuum, the CFT path integral on this strip implements a finite linear map $V = e^{-Q}$ generated by $Q = Q_0 + i Q_1$ where $Q_0$ (non-uniform euclidean time evolution) dominates and $Q_1$ (space diffeomorphism) is only a small correction \cite{SupplMat}. On the lattice, the tensor networks act on the low energy states of a sequence of periodic quantum spin chain of size $N = 16,32,64$, where $\aUV = 2\pi/N$. As we decrease $\aUV$, the discrete profile of euclideons approximates better the continuous function $f(x)$, suggesting a better approximation to the continuous strip geometry. Indeed, the linear map implemented by the tensor network is seen to better approximate the conformal map $V = e^{-Q}$ obtained through the path integral in the continuum \cite{SupplMat}.

\textit{Discussion.---} In this \letter we have proposed how to assign geometries to a class of tensor networks describing quantum critical spin chains. Our main contribution is rule 1 above, which requires that the linear map implemented by a strip of the network should be compatible with the linear map implemented in the continuum by the CFT path integral on a strip of the assigned geometry. The second requirement, namely that the lattice spacing correspond to a constant distance $\aUV$ throughout the network (rule 2 above), has also been used in most previous proposals to assign geometries to tensor networks \cite{H2,dS2,EHM,HAPPY,Random,PathIntegral1,PathIntegral2,TNRlocal}. Several remarks are in order. (\textit{i}) While rule 2 conveniently allowed us to assign a single length scale $\aUV$ throughout the tensor network, the tensor network formalism can also be used to describe a CFT on a lattice with inhomogeneous lattice spacing $\aUV(\tau,x)$, in which case rule 2 is replaced with the choice of scale factor $\Omega(\tau,x) \sim \aUV(\tau,x)$, see e.g. Refs. \cite{Quotient, QFTCS}. (ii) By considering lorentzions $l$ (such that a layer of lorentzions $l$ implements a map $e^{-iH}$) instead of euclideons $e$, our present proposal extends also to 2d spacetime geometries with lorentzian signature \cite{QFTCS}. (iii) Finally, a similar geometric assignment is possible for tensor networks describing lattice models whose low energy physics correspond instead to a massive quantum field theory. In this case the tensors in the network (euclideons, disentanglers and isometries) are no longer scale-invariant but seen to depend explicitly on the UV cut-off scale $\aUV(\tau,x)$ \cite{QFTCS}.
 
A main motivation for this work was to develop natural criteria to assign a geometry to the MERA. As discussed in Ref.\ \cite{geoMERA}, from the perspective of the path integral advocated here, MERA corresponds to a degenerate geometry (the metric has zero determinant) which is therefore neither the hyperbolic plane H$_2$ \cite{H2} nor de Sitter spacetime dS$_2$ \cite{dS2}. However, the MERA tensor network is also shown to admit euclidean and lorentzian extensions that correspond to a path integral in both H$_2$ and dS$_{1,1}$, thus realizing the scenarios described in Refs. \cite{H2} and \cite{dS2}. 


\textit{Acknowledgments.} We thank 
Bartlomiej Czech,
Pawel Caputa,
William Donnelly,
Davide Gaiotto,
Qi Hu,
Lampros Lamprou,
Juan Maldacena,
David Mateos,
Samuel McCandlish,
Rob Myers,
James Sully,
Vasudev Shyam, 
Tadashi Takayanagi,
and Xiao-liang Qi
for fruitful discussions and feedback.
The authors acknowledge
support by the Simons Foundation (Many Electron
Collaboration), by NSERC (discovery grant), and by Compute Canada. Research at
Research at Perimeter Institute is supported by the Government of Canada through the Department of Innovation, Science and Economic Development Canada and by the Province of Ontario through the Ministry of Research, Innovation and Science.

\section{Appendix: 2d geometry as linear maps between time slices}
\label{sec:AppI}

Sections \ref{subsec:1A} to \ref{subsec:1I} we provide a detailed derivation of Eqs.\ 1-6 of the main text. This is standard material that can be found scattered in most graduate-level QFT/CFT books. We compiled this material together in uniform notation for completeness. These sections do not contain original research.

\subsection{Slicing}
\label{subsec:1A}

We consider a 2d euclidean spacetime $(\mathcal{M},g)$ where, for simplicity, the manifold $\mathcal{M}$ can be decomposed into non-overlapping, one-dimensional slices $\Sigma_\tau$
\begin{equation} \label{eq:MSigma}
\mathcal{M} = \bigcup_{\tau \in I} \Sigma_{\tau},~~~\Sigma_\tau \cap \Sigma_{\tau'} = \emptyset,
\end{equation}
and where each slice $\Sigma_\tau$ is topologically equivalent to a fixed one-dimensional manifold $\Sigma$, so that topologically we have $\mathcal{M} = I \times \Sigma$. We will think of $\Sigma$ as a space-like surface and of $\tau$ as labeling euclidean time. A concrete example would be a topological cylinder, where the space-like surface $\Sigma$ is a circle S$_1$. 

We choose coordinates $x^{\mu} = (\tau,x)$ on $\mathcal{M}$ such that, by construction, the euclidean time $\tau$ is constant on each slice $\Sigma_\tau$. In these coordinates the metric $g$ reads
\begin{eqnarray}\label{eq:g1}
g_{\mu\nu}(\tau,x) &=& \left(
\begin{array}{cc}
g_{00}(\tau,x) & g_{01}(\tau,x) \\
g_{10}(\tau,x) & g_{11}(\tau,x)
\end{array} \right)\\
&=&\Omega^{2}(\tau,x)\left(\begin{array}{cc} A(\tau,x)^2 & B(\tau,x)\\ B(\tau,x) & 1 \end{array} \right).\label{eq:g2}
\end{eqnarray} 
Here we have explicitly extracted a scale factor $\Omega^{2}(\tau,x)$ for later convenience. Recall that the euclidean metric is positive defined, so that the determinant $|g| \equiv \det(g) = \Omega^4(A^2-B^2)$ is positive, that is $A^2 > B^2$ for all $(\tau,x)$. The inverse metric reads
\begin{eqnarray}
g^{\mu\nu}(\tau,x) &=& \frac{1}{|g|}\left(
\begin{array}{cc}
g_{11} & -g_{01} \\
-g_{10} & g_{00}
\end{array} \right)\\
&=& \frac{1}{\Omega^2(A^2-B^2)}\left(\begin{array}{cc} 1 & -B\\ -B & A^2 \end{array} \right).
\end{eqnarray} 

In the tangent space of $\mathcal{M}$ at $\Sigma_\tau$, let
\begin{equation} \label{eq:xi1}
\xi^{\mu}(\tau,x) = (1,0) 
\end{equation}
be the vector such that, for small $\epsilon \ll 1$, $\epsilon \xi^{\mu}$ maps each point $p^{\mu}\equiv(\tau,x)$ in slice $\Sigma_{\tau}$ into a corresponding point $q^{\mu}\equiv(\tau+\epsilon,x)$ in slice $\Sigma_{\tau+\epsilon}$ (that is, without changing the coordinate $x$),
\begin{equation}
q^{\mu} = p^{\mu} + \epsilon \xi^{\mu},
\end{equation}
and let $n^{\mu}(\tau,x)$ be the unit normal vector to $\Sigma_{\tau}$,
\begin{eqnarray}  \label{eq:n1}
n_{\mu} &\equiv& N \partial_{\mu} \tau = N (1,0) \\
&=& \sqrt{\frac{g}{g_{11}}}(1,0) = \Omega\sqrt{A^2-B^2}(1,0),
\end{eqnarray}
or
\begin{eqnarray}
n^{\mu} &=& g^{\mu\nu}n_{\nu} = \frac{1}{\sqrt{gg_{11}}}(g_{11},-g_{1 0})\\
&=& \frac{1}{\Omega\sqrt{A^2-B^2}}(1, -B),\label{eq:n4}
\end{eqnarray}
where the normalization $N$ above was fixed by the condition $n^{\mu}n_{\mu}=1$. We will be interested in the scalar products $\xi^{\mu}\xi_{\mu}$ and $\xi^{\mu}n_{\mu}$, which read
\begin{eqnarray} \label{eq:xixi}
\xi^{\mu}\xi_{\mu} &=& g_{\mu\nu}\xi^{\mu}\xi^{\nu} = g_{00}=\Omega^2 A^2,\\
\xi^{\mu}n_{\mu} &=& \sqrt{\frac{g}{g_{11}}}=\Omega\sqrt{A^2-B^2}. \label{eq:xin}
\end{eqnarray}

\begin{figure}
\includegraphics[width=\linewidth]{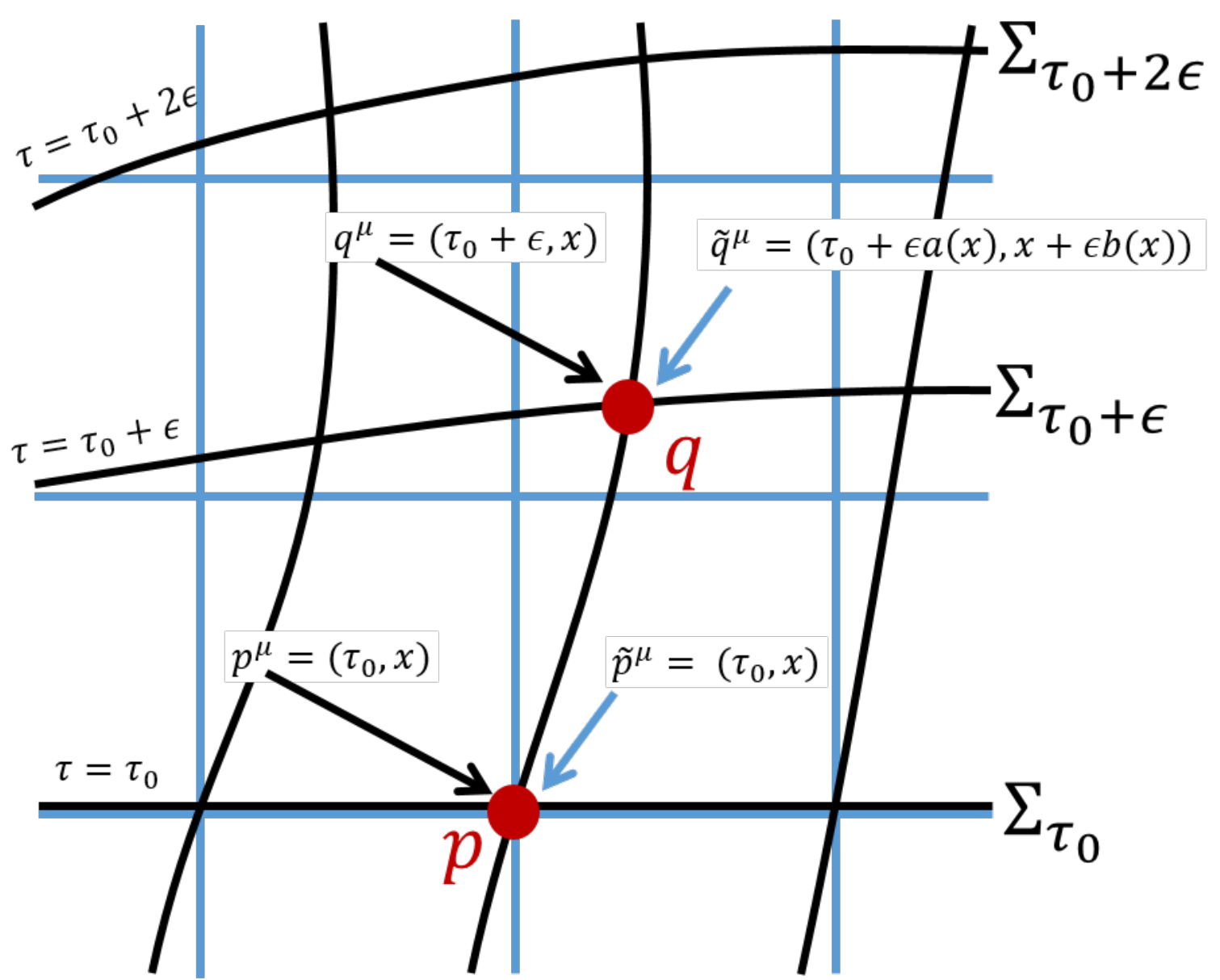}
\caption{\label{fig:coordinates} 
Points $p\in\Sigma_{\tau_0}$ and $q\in\Sigma_{\tau_0+\epsilon}$ expressed both in the $x^{\mu}=(\tau,x)$ coordinates (in black) and in auxiliary conformal coordinates $y^{\mu} = (y^{0},y^{1})$ (in blue). 
}
\end{figure}

\subsection{Auxiliary conformal coordinates}
\label{subsec:1B}

\begin{figure}
\includegraphics[width=7cm]{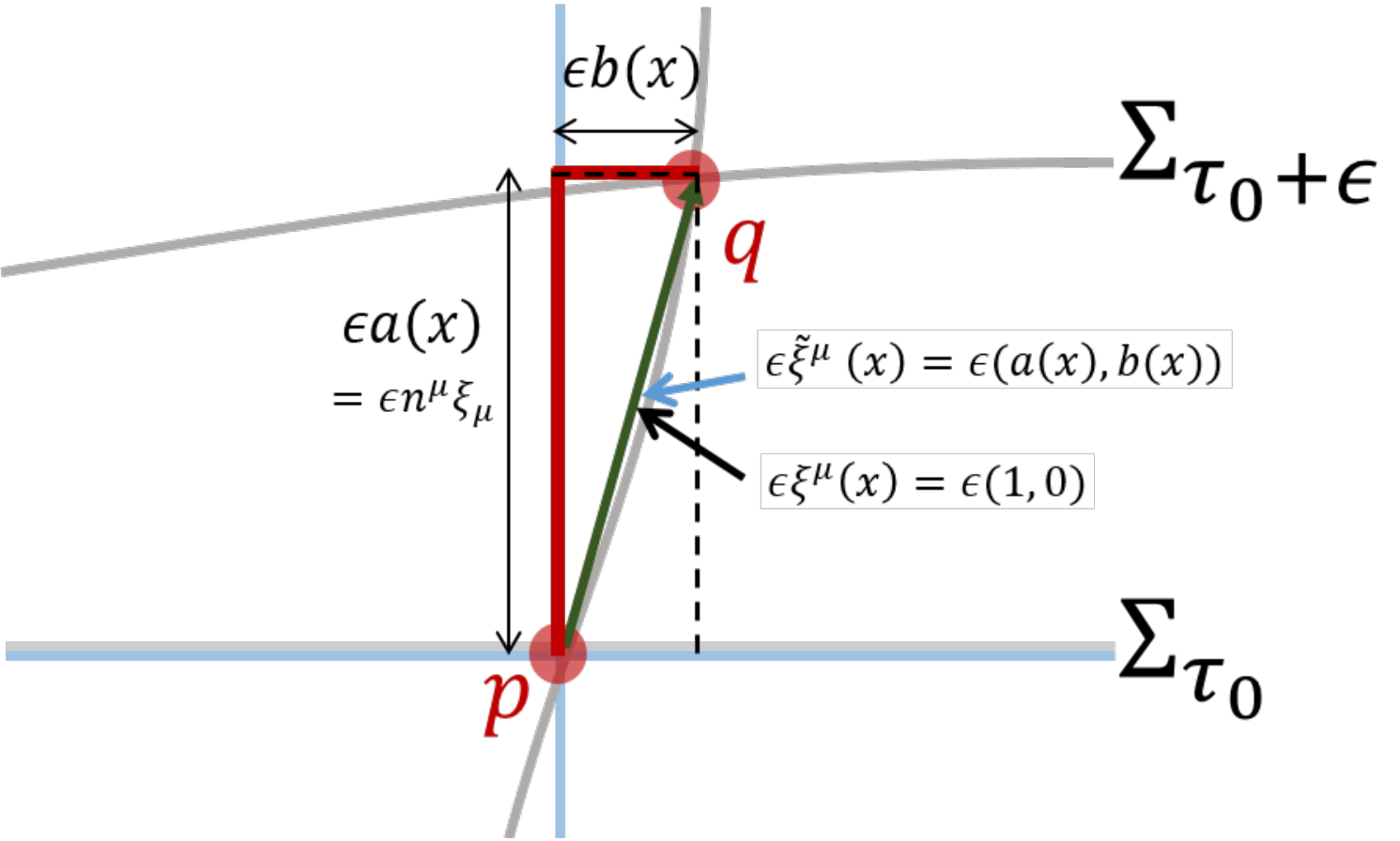}
\caption{\label{fig:lapseshift} 
The transition vector $\xi^{\mu}$ between slice $\Sigma_{\tau_0}$ and $\Sigma_{\tau_0+\epsilon}$ expressed in coordinates $x^{\mu}=(\tau,x)$, namely $\xi^{\mu}(x) = (1,0)$, and auxiliary conformal coordinates $y^{\mu} = (y^0,y^1)$, namely $\tilde{\xi}^{\mu}(x) = (a(x), b(x))$. The so-called lapse $\alpha(x) \equiv \epsilon a(x)$ and shift $\beta(x) \equiv \epsilon b(x)$ are, respectively, the projections of $\epsilon \xi^{\mu}$ onto the unit vector $n^{\mu}$ normal to the slice $\Sigma_{\tau_0}$ and the projections of $\epsilon \xi^{\mu}$ onto the time slice $\Sigma_{\tau_0}$ itself. 
}
\end{figure}

Next we introduce an auxiliary set of conformal coordinates $(y^{0},y^{1})$ such that they coincide with $(\tau,x)$ on the time slice $\Sigma_{\tau_0}$, that is, $(y_0,y_1) = (\tau,x)$ when $\tau = \tau_0$, see Fig.\ \ref{fig:coordinates}. As in any set of conformal coordinates, the metric $g$ expressed in the coordinates $(y^{0},y^{1})$ is proportional to the identity, 
\begin{equation}
~~~\tilde{g}_{\mu\nu}(y^{0},y^{1}) = \tilde{\Omega}^2(y^{0},y^{1}) \left(
\begin{array}{cc}
1 & 0 \\
0 & 1
\end{array}
\right),
\end{equation}
and
\begin{equation}
~~~\tilde{g}^{\mu\nu}(y^{0},y^{1}) = \frac{1}{\tilde{\Omega}^2} \left(
\begin{array}{cc}
1 & 0 \\
0 & 1
\end{array}
\right).
\end{equation}
Moreover, by construction the scale factor $\tilde{\Omega}$ coincides with $\Omega$ on the slice $\Sigma_{\tau_0}$,
\begin{equation}
\tilde{\Omega}(\tau_0,x) = \Omega(\tau_0,x) ~~~~\forall x. 
\end{equation}
Let 
\begin{equation} \label{eq:xi2}
\tilde{\xi}^{\mu}(y^0,y^1) = (a(y^0,y^1),b(y^0,y^1))
\end{equation}
be the transition vector $\xi^{\mu}$ between slices $\Sigma_{\tau_0}$ and $\Sigma_{\tau_0+\epsilon}$, Eq.~\eqref{eq:xi1}, when expressed in the conformal coordinates $(y^0,y^1)$, see Figs. \ref{fig:coordinates}-\ref{fig:lapseshift}. In these coordinates the unit normal vector $n^{\mu}$ in Eq.~\eqref{eq:n1}-\eqref{eq:n4} now reads
\begin{equation}
\tilde{n}^{\mu} = \frac{1}{\tilde{\Omega}}(1,0),~~\tilde{n}_{\mu} = \tilde{\Omega}(1,0). 
\end{equation}
The scalar products $\xi^{\mu}\xi_{\mu}$ and $\xi^{\mu}n_{\mu}$ in Eq.~\eqref{eq:xixi}-\eqref{eq:xin}, which are invariant under a change of coordinates, now read
\begin{eqnarray}\label{eq:xixi2}
\xi^{\mu}\xi_{\mu} &=& \tilde{\xi}^{\mu}\tilde{\xi}_{\mu}  = \Omega^2 (a^2+b^2),\\
\xi^{\mu}n_{\mu} &=& \tilde{\xi}^{\mu}\tilde{n}_{\mu} = \Omega a.\label{eq:xin2}
\end{eqnarray}
We thus conclude, comparing Eqs.\ \eqref{eq:xixi}-\eqref{eq:xin} with Eqs.\ \eqref{eq:xixi2}-\eqref{eq:xin2}, that
\begin{equation} \label{eq:compare1}
a = \sqrt{A^2-B^2},~~~~b = B,
\end{equation}
or, equivalently,
\begin{equation} \label{eq:compare2}
A = \sqrt{a^2+b^2},~~~~B = b.
\end{equation}

\subsection{Path integral and transition amplitudes}
\label{subsec:1C}
 
On the euclidean spacetime ($\mathcal{M},g$) parameterized by $\mathbf{x} = x^{\mu} = (\tau,x)$ we define a local QFT, characterized by a field $\phi(\mathbf{x})$ (possibly representing a set of several fields) and the path integral
\begin{equation} \label{eq:path1}
\mathcal{Z} = \int [D\phi]~ e^{-S[\phi]},~~~~~ 
\end{equation}
where $\int [D\phi]$ is an integral over field configurations and $S[\phi]$ is the action of a given field configuration $\phi$, that is
\begin{equation}
S[\phi] \equiv \int_{\mathcal{M}}\!\!\! d\tau\, dx ~\sqrt{|g|} ~\mathcal{L}\left(\phi(\mathbf{x}), \partial_{\mu}\phi(\mathbf{x})\right),
\end{equation}
where $\mathcal{L}\left(\phi(\mathbf{x}), \partial_{\mu}\phi(\mathbf{x})\right)$ is a local lagrangian density.

Given two times $\tau_{\inn}$ and $\tau_{\out}$, with $\tau_{\out}>\tau_{\inn}$, let $\mathcal{N}$ denote the strip of geometry that has the corresponding slices $\Sigma_{\inn}$ and $\Sigma_{\out}$ as its boundaries. Then the path integral on $\mathcal{N}$ with the field boundary conditions
\begin{equation}
\phi(\tau=\tau_{\inn},x) = \varphi(x),~~~~ \phi(\tau=\tau_{\out},x) = \varphi'(x)
\end{equation}
defines a transition amplitude
\begin{eqnarray} \label{eq:Aphiphi}
A(\varphi(x) \rightarrow \varphi'(x)) \equiv \int [D\phi]~ e^{-S_{\mathcal{N}}[\phi]}.
\end{eqnarray}

\subsection{Hilbert space and linear maps}
\label{subsec:1D}

Given a slice $\Sigma_\tau$, we define the Hilbert space $\mathcal{H}(\Sigma_\tau)$ in terms of a basis of vectors $\ket{\varphi(x)}$, where each such vector corresponds to a field configuration $\varphi(x)$ on $\Sigma_\tau$. Further, given the two slices $\Sigma_{\inn}$ and $\Sigma_{\out}$, we identify basis vectors on the corresponding Hilbert spaces $\mathcal{H}(\Sigma_{\inn})$ and $\mathcal{H}(\Sigma_{\out})$ through
\begin{equation} \label{eq:identification}
 \ket{\varphi(x)}_{\Sigma_{\inn}} \sim \ket{\varphi'(x)}_{\Sigma_{\out}},~\mbox{iff}~~\varphi(x)=\varphi'(x)~\mbox{for all}~x. 
 \end{equation} 
This identification between states in the Hilbert space, which depends on the choice of spatial coordinate $x$ on $\Sigma_{\inn}$ and $\Sigma_{\out}$, allows us to consider a single Hilbert space $\mathcal{H}(\Sigma)$ for all slices $\Sigma_{\tau}$.
 
We then use the transition amplitude $A(\varphi(x) \rightarrow \varphi'(x))$ in Eq.~\eqref{eq:Aphiphi} to define a linear map $V:\mathcal{H}(\Sigma_{\inn})\rightarrow\mathcal{H}(\Sigma_{\out})$ with matrix elements
\begin{eqnarray}
\bra{\varphi'(x)} V \ket{\varphi(x)} = \int [D\phi]~ e^{-S_{\mathcal{N}}[\phi]}.
\end{eqnarray}
Through the above identification, it becomes a linear map $V:\mathcal{H}(\Sigma)\rightarrow\mathcal{H}(\Sigma)$. 

\subsection{Generator of the linear map for a thin strip}
\label{subsec:1E}

When $\tau_{\out} = \tau_{\inn}+\epsilon$ for a small $\epsilon \ll 1$, $\mathcal{N}$ is a thin strip and we can expand the linear map $V$ to linear order in $\epsilon$,  
\begin{equation} \label{eq:V2}
V = e^{-\epsilon Q} \approx \mathbb{I} - \epsilon Q.
\end{equation}
As reviewed in the derivation below, $Q$ can be written in terms of the euclidean stress tensor $T_{E}$ as
\begin{eqnarray}
Q &=& -\int_{\Sigma_{\inn}}\!\! dx~\sqrt{g_{11}(\mathbf{x})}  ~n^{\mu}(\mathbf{x})~ \xi^{\nu}(\mathbf{x})~ (T_E)_{\mu\nu}(\mathbf{x}).~~~~~ \nonumber
\end{eqnarray}
Here $dx\sqrt{g_{11}(\mathbf{x})}$ is the proper length attached to an infinitesimal $dx$ of the space coordinate $x$, $n^{\mu}(\mathbf{x})$ is the unit normal vector of $\Sigma_{\inn}$ at position $x$, and $\xi^{\mu}(\mathbf{x})$ is the transition vector such that $\epsilon \xi^{\mu}(\mathbf{x})$ maps points in $\Sigma_{\inn}$ to points in $\Sigma_{\out}$ without changing the value of the space coordinate $x$. Given that both $dx\sqrt{g_{11}(\mathbf{x})}$ and $n^{\mu}(\mathbf{x})\xi^{\nu}(\mathbf{x})T_{\mu\nu}(\mathbf{x})$ transform as scalars under a change of coordinates, we can evaluate their product in the auxiliary conformal coordinates $(y^{0},y^{1})$, which are adjusted to coincide with $(\tau,x)$ at $\Sigma_{\inn}$, that is, such that $(y^{0},y^{1})=(\tau,x)$ for $\tau=\tau_{\inn}$. We then find
\begin{eqnarray}
&&dx \sqrt{g_{11}}n^{\mu}\xi^{\nu}(T_E)_{\mu\nu} \\
&=& dx \sqrt{\tilde{g}_{11}}\tilde{n}^{\mu}\tilde{\xi}^{\nu}(\tilde{T}_E)_{\mu\nu} \\
&=& dx \left(a(\mathbf{x}) (\tilde{T}_E)_{00}(\mathbf{x}) + b(\mathbf{x})(\tilde{T}_E)_{01}(\mathbf{x})\right),~~~~~
\end{eqnarray}
where $a(\mathbf{x})$ and $b(\mathbf{x})$ are the components of $\tilde{\xi}^{\mu}(\mathbf{x})$ in Eq.~\eqref{eq:xi2}, and in these expressions $\mathbf{x} = (\tau_{\inn},x)$. 

We proceed by specializing to a CFT. In conformal coordinates, the stress tensor always has the same components $(\tilde{T_E)_{00}} = h_E$ and $(\tilde{T}_E)=p_E$ (up to an additive constant due to the conformal anomaly, which for the purposes of this work we can ignore, see \cite{Weyl}), where $h_E$ and $p_E$ are the euclidean energy and momentum densities. In terms of the more conventional lorentzian energy and momentum densities $h = -h_E$ and $p = ip_E$ (see free boson example below), we arrive to
\begin{eqnarray} \label{eq:QQ}
Q &=&\int_{\Sigma_{\inn}} dx\left(a(\mathbf{x}) h(\mathbf{x}) + i b(\mathbf{x}) p(\mathbf{x}) \right)\\
&=& Q_0 + i Q_1,
\end{eqnarray}
where we have split the generator $Q = Q_0 + iQ_1$ into two contributions 
\begin{eqnarray}
Q_0 &\equiv& \int_{\Sigma_{\inn}} dx~a(\mathbf{x}) ~h(\mathbf{x}), \\
Q_1 &\equiv& \int_{\Sigma_{\inn}} dx~b(\mathbf{x}) ~p(\mathbf{x}), 
\end{eqnarray}
that generate non-uniform time and space translations, respectively.

\subsection{Stress tensor and Ward identity}
\label{subsec:1F}
\label{subsec:1G}

In our derivation below, the euclidean stress tensor $(T_E)_{\mu\nu}$ plays an important role. It can be defined in two alternative ways. First, using Noether's theorem, as a pair of classically conserved currents (corresponding to the invariance of the action $S[\phi(\mathbf{x})]$ under time and space translations). Second, as characterizing the variations of the action $S[\phi]$ under changes of the metric.

Consider an infinitesimal change of coordinates of the form
\begin{equation} \label{eq:Axx}
x^{\mu} \rightarrow x'^{\mu} = x^{\mu} + \epsilon^{\mu}(\mathbf{x}),
\end{equation}
and a new field $\phi'(\mathbf{x}')$ through
\begin{equation} 
\phi'(\mathbf{x}') \equiv \phi(\mathbf{x}).
\end{equation}
We notice that
\begin{eqnarray}
 \frac{\partial x'^{\mu}}{\partial x^{\nu}} = \delta^{\mu}_{\nu} +\frac{\partial \epsilon^{\mu}}{\partial x^{\nu}}, ~~~~~
 \frac{\partial x^{\mu}}{\partial x'^{\nu}} = \delta^{\mu}_{\nu} -\frac{\partial \epsilon^{\mu}}{\partial x'^{\nu}},
\end{eqnarray}
and therefore the determinant of the Jacobian reads
\begin{equation}
\left|\frac{\partial x}{\partial x'}\right| = 1 - \partial_{\mu}\epsilon^{\mu},
\end{equation}
whereas 
\begin{eqnarray}
\partial_{\mu} \phi(\mathbf{x})&\equiv&\frac{\partial \phi(\mathbf{x})}{\partial x^{\mu}} 
= \frac{\partial x'^{\rho}}{\partial x^{\mu}} \frac{\partial \phi'(\mathbf{x}')}{\partial x'^{\rho}} \\
&=&  \left(\delta^{\rho}_{\mu} +\frac{\partial \epsilon^{\rho}}{\partial x^{\mu}}\right) \frac{\partial \phi'(\mathbf{x}')}{\partial x'^{\rho}} \\
&=& \frac{\partial \phi'(\mathbf{x}')}{\partial x'^{\mu}} + \frac{\partial \epsilon^{\rho}}{\partial x^{\mu}} \frac{\partial \phi'(\mathbf{x}')}{\partial x'^{\rho}}\\
&=& \partial'_{\mu} \phi'(\mathbf{x}') + (\partial'_{\mu}\epsilon^{\rho}(\mathbf{x}')) \partial'_{\rho} \phi'(\mathbf{x}').
\end{eqnarray}

Given a field configuration $\phi(\mathbf{x})$, let us rewrite its action $S[\phi(\mathbf{x})]$ as
\begin{eqnarray}
 &&S[\phi(\mathbf{x})] \equiv \int d\tau dx~\mathcal{L}(\phi(\mathbf{x}), \partial_{\mu}\phi(\mathbf{x}))\label{eq:AS1} \\ 
&=& \int d\tau' dx' ~\left(1-\partial_{\mu}\epsilon^{\mu}(\mathbf{x}') \right) ~\times \label{eq:AS2}\\
&& \mathcal{L}\left(\phi'(\mathbf{x}'), \partial'_{\mu} \phi'(\mathbf{x}') + (\partial'_{\mu}\epsilon^{\nu}(\mathbf{x}'))\partial'_{\nu}\phi'(\mathbf{x}')\right)~~~~\nonumber\\
&=& \int d\tau dx ~\left(1-\partial_{\mu}\epsilon^{\mu}(\mathbf{x})\right)~\times \label{eq:AS3} \\
&& \mathcal{L}\big(\phi'(\mathbf{x}), \partial_{\mu} \phi'(\mathbf{x}) + (\partial_{\mu}\epsilon^{\nu}(\mathbf{x})) \partial_{\nu} \phi'(\mathbf{x})\big)~~~~ \nonumber \\
&=& \int d\tau dx ~\left(1-\partial_{\mu}\epsilon^{\mu}(\mathbf{x})\right)~\times \label{eq:AS4}\\
&&\left(  \mathcal{L}\left(\phi'(\mathbf{x}), \partial_{\mu} \phi'(\mathbf{x})\right)  + \frac{\delta \mathcal{L}}{\delta \partial_\mu \phi' } (\partial_{\mu}\epsilon^{\nu}(\mathbf{x})) \partial_{\nu} \phi'(\mathbf{x}) \right)~~~ \nonumber\\
&=&\int d\tau  dx~\mathcal{L}\left(\phi'(\mathbf{x}), \partial_{\mu} \phi'(\mathbf{x})\right) \label{eq:AS5}\\
 &+& \int d\tau dx~\left(\partial_{\mu}\epsilon^{\nu}(\mathbf{x})\right) \left( - \delta_{\nu}^{\mu} \mathcal{L} + \frac{\delta \mathcal{L}}{\delta \partial_\mu \phi'}\partial_{\nu} \phi'(\mathbf{x})\right) \nonumber\\
&=& S[\phi'(\mathbf{x})] ~+ ~\int d\tau dx~\left(\partial_{\mu}\epsilon^{\nu}(\mathbf{x})\right) (T_{E})_{\nu}^{\mu} (\mathbf{x}) \label{eq:AS6}\\
&=& S[\phi'(\mathbf{x})] ~- ~\int d\tau dx~\left(\partial_{\mu}(T_{E})_{\nu}^{\mu} (\mathbf{x})\right) \epsilon^{\nu}(\mathbf{x}) \label{eq:AS7}
\end{eqnarray}
where \eqref{eq:AS1} is the definition, in \eqref{eq:AS2} we changed coordinates, in \eqref{eq:AS3} we simply renamed the coordinates $x'^{\mu}$ as $x^{\mu}$ (since they are integration variables), in \eqref{eq:AS4} we expanded the Lagrangian according to
\begin{eqnarray}
\mathcal{L}\left(\phi', \partial_{\mu} \phi' + \delta A \right) = \mathcal{L}\left(\phi', \partial_{\mu} \phi'\right) + \frac{\delta \mathcal{L}\left(\phi', \partial_{\mu} \phi'\right)}{\delta \partial_\mu \phi' } \delta A, ~~~~~\label{eq:parts}
\end{eqnarray}
in \eqref{eq:AS5} we separated a first term independent of $\epsilon^{\mu}(\mathbf{x})$ and a second term that collects all contributions linear in $\partial_{\mu} \epsilon^{\nu}(\mathbf{x})$ (the contribution linear in $\epsilon^{\nu}$ must vanish since the action is invariant under uniform time and space translations), in \eqref{eq:AS6} we identified the first term as the original action evaluated on the new field configuration $\phi'(\mathbf{x}) = \phi(\mathbf{x-\epsilon})$ and the second term as an integral of $\partial_{\mu} \epsilon^{\nu}(\mathbf{x})$ times the euclidean stress tensor
\begin{equation}
(T_E)^{\mu}_{\nu}(\mathbf{x}) = - \delta^{\mu}_{\nu}\mathcal{L} + \frac{\delta \mathcal{L}}{\delta \partial_{\mu}\phi} \partial_{\nu}\phi,
\end{equation}
evaluated on the field configuration $\phi'(\mathbf{x})$ (or on the field $\phi(\mathbf{x})$, up to corrections that are higher order in $\epsilon^{\mu}(\mathbf{x})$) and, finally, in \eqref{eq:AS7} we used integration by parts in the second term. Noether's theorem applied to invariance of the action under uniform time-space translations implies that the stress tensor is conserved
\begin{equation}
\partial_{\mu} (T_E)_{\nu}^{\mu} (\mathbf{x}) = 0,
\end{equation}
when evaluated on classical field configurations $\phi_{cl}(\mathbf{x})$, that is, field configurations that extremize the action functional (and thus obey the classical equations of motion).

Alternatively, we could first rewrite the action functional $S[\phi(\mathbf{x})]$ in an explicitly covariant form by introducing the metric $g_{\mu\nu}(\mathbf{x})$,
\begin{eqnarray}
S[\phi(\mathbf{x}), g_{\mu\nu}(\mathbf{x})] = \int d\tau dx \sqrt{|g|}~\mathcal{L}(\phi(\mathbf{x}), \partial_{\mu}(\mathbf{x}), g(\mathbf{x})), ~~~~~~
\end{eqnarray}
and then define the euclidean stress tensor $(T_E)_{\mu\nu}$ in terms of the variations of the action under changes of the metric (which it is then set back to being flat, that is $g_{\mu\nu} = \delta_{\mu\nu}$)
\begin{equation}
(T_E)_{\mu\nu}(\mathbf{x}) \equiv -\left.\frac{2}{\sqrt{|g|}}\frac{\delta S}{\delta g^{\mu\nu}(\mathbf{x})}\right|_{g_{\rho\sigma}(\mathbf{x}) = \delta_{\rho\sigma}}.
\end{equation}
see below the case of a free boson as an example.

Let us now consider the path integral with 
\begin{equation} \label{eq:API1}
\mathcal{Z}_{\mathcal{O}} = \int [D\phi]~ \mathcal{O} ~e^{-S[\phi]},~~~~~ 
\end{equation}
where $S[\phi(\mathcal{x})]$ is the action functional of Eq.\ \eqref{eq:AS1} and $\mathcal{O}$ could be a single local insertion $\mathcal{O}_1(\mathbf{x}_1)$, e.g. $\mathcal{O}_{1}(\mathbf{x}_1) = \phi(\mathbf{x}_1)^{2}\partial^{\mu}\partial_{\mu}\phi(\mathbf{x}_1)$ or, more generally, a sequence of such local insertions $\mathcal{O}_1(\mathbf{x}_1)\mathcal{O}_2(\mathbf{x}_2) \cdots \mathcal{O}_n(\mathbf{x}_n)$ at points $\mathbf{x}_1, \mathbf{x}_2, \cdots, \mathbf{x}_n$.
Let us rewrite this expression considering the above infinitesimal change of coordinates \eqref{eq:Axx} were we will assume for simplicity that the coordinates do not change (that is, $\epsilon^{\mu}(\mathbf{x})$ vanishes) in a finite neighborhood of the location $\mathbf{x}_i$ of each insertion in $\mathcal{O}$. Then we have
\begin{eqnarray}
&&\mathcal{Z}_{\mathcal{O}} = \int [D\phi]~ \mathcal{O} ~e^{-S[\phi]} \label{eq:AW1}\\
&=& \int [D\phi'] \mathcal{O} e^{-S[\phi'] + \int d\tau dx ~\partial_{\mu}(T_E)_{\nu}^{\mu}(\mathbf{x}) \epsilon^{\nu}(\mathbf{x})} \label{eq:AW2}\\
&=&  \int [D\phi'] \mathcal{O} e^{-S[\phi']} \label{eq:AW3}\\
&+&  \int [D\phi'] \mathcal{O} e^{-S[\phi']} \left(
\int d\tau dx ~\partial_{\mu}(T_E)_{\nu}^{\mu}(\mathbf{x}) \epsilon^{\nu}(\mathbf{x}) \right). \nonumber
\end{eqnarray}
In \eqref{eq:AW2} we assumed as customary that the functional integration measure $[D\phi]$ is invariant under the field transformation, that is $[D\phi'] = [D\phi]$, and used Eqs.\ \eqref{eq:AS1}-\eqref{eq:AS7}. Then in \eqref{eq:AW3} we separate a first term independent of $\epsilon^{\mu}(\mathbf{x})$ from a second term that collects all linear order contributions in $\epsilon^{\mu}(\mathbf{x})$. Now we notice that the field $\phi'(\mathbf{x})$ is an integration variable and we can change its name to $\phi(\mathcal{x})$, which makes manifest that this first term is equal to $\mathcal{Z}_{\mathcal{O}}$. Therefore we arrive to
\begin{equation}
\int d\tau dx  ~\epsilon^{\nu}(\mathbf{x}) ~ \left(\int [D\phi'] \mathcal{O} e^{-S[\phi']} 
\partial_{\mu}(T_E)_{\nu}^{\mu}(\mathbf{x}) \right) = 0,
\end{equation}
which is valid for all $\epsilon^{\mu}(\mathbf{x})$ (with the above restriction) and thus implies that
\begin{equation}
\int [D\phi'] ~\mathcal{O}~ e^{-S[\phi']} 
~\partial_{\mu}(T_E)_{\nu}^{\mu}(\mathbf{x}) = 0 \label{eq:Ward}
\end{equation}
for all $\mathbf{x}$ which does not coincide with the insertion(s) $\mathcal{O}$, that is $\mathbf{x} \not = \mathbf{x}_1, \mathbf{x}_2, \cdots \mathbf{x}_n$.

\subsection{Infinitesimal evolution of the wavefunctional}
\label{subsec:1H}

\begin{figure}
\includegraphics[width=\linewidth]{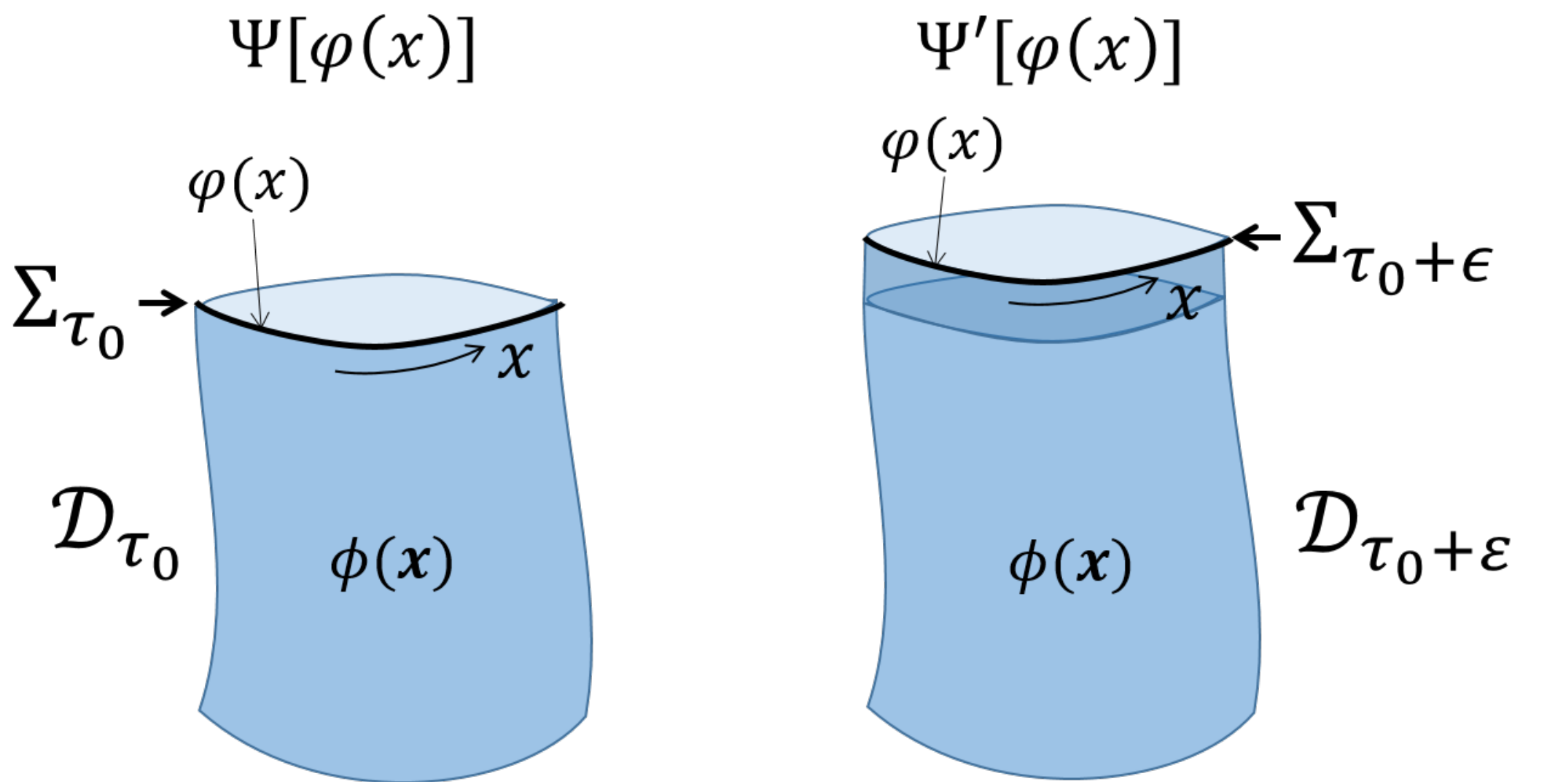}
\caption{\label{fig:wavefunctionals} 
(Left) The wavefunctional $\Psi[\varphi(x)]$ is obtained through a path integral on region $\mathcal{D}_{\tau_0}$, where the field configurations $\phi(\mathbf{x})$ have the boundary condition $\phi(\tau_0,x) = \varphi(x)$ on the time slice $\Sigma_{\tau_0}$.
(Left) The evolved wavefunctional $\Psi'[\varphi(x)]$ is obtained through a path integral on region $\mathcal{D}_{\tau_0+\epsilon}$, where the field configurations $\phi(\mathbf{x})$ now have the boundary condition $\phi(\tau_0+\epsilon,x) = \varphi(x)$ on the time slice $\Sigma_{\tau_0+\epsilon}$.
}
\end{figure}

Consider the wavefunctional $\Psi[\varphi(x)]$ on the time slice $\Sigma_{\tau_0}$ at time $\tau=\tau_0$, defined through a path integral on the domain $\mathcal{D}_{\tau_0}\subseteq \mathcal{M}$, consisting of all times $\tau\leq \tau_0$, as
\begin{eqnarray}
\Psi[\varphi(x)] &\equiv& \int [D\phi]~ \mathcal{O} ~e^{-S^{(\tau_0)}[\phi(\mathbf{x})]}, \\
&&\phi(\mathbf{x}\in \mathcal{D}_{\tau_0}), \\
&&\phi\left(\tau=\tau_0,x\right) = \varphi(x).
\end{eqnarray}
Here the path integral is over field configurations $\phi(\mathbf{x})$ for $\mathbf{x}\in \mathcal{D}_{\tau_0}$ and such that they coincide with $\varphi(x)$ at the boundary $\Sigma_{\tau_0}$ of $\mathcal{D}_{\tau_0}$, that is $\phi(\tau_0,x) = \varphi(x)$. The action functional now reads
\begin{eqnarray}
S^{(\tau_0)}[\phi(\mathbf{x})] &\equiv&\int_{-\infty}^{\tau_0} d\tau \int dx~\mathcal{L}(\phi(\mathbf{x}), \partial_{\mu}\phi(\mathbf{x})),
\end{eqnarray}
By changing the insertion(s) $\mathcal{O}$, we could obtain a different wavefunctional $\Psi[\varphi(x)]$. Notice that the wavefunctional $\Psi[\varphi(x)]$ corresponds to a vector $\ket{\Psi}$ in the Hilbert space $\mathcal{H}(\Sigma_{\tau_0})$, 
\begin{equation}
\ket{\Psi} \in \mathcal{H}(\Sigma_{\tau_0}), ~~~\Psi[\varphi(x)] = \braket{\varphi(x)}{\Psi}.
\end{equation}

Our goal is to evolve the wavefunctional $\Psi[\varphi(x)]$ from time $\tau = \tau_0$ to time $\tau = \tau_0 + \epsilon$ using the path integral in the intermediate strip, which will produce a new wavefunctional $\Psi'[\varphi(x)]$
\begin{eqnarray}
\Psi'[\varphi(x)] &\equiv& \int [D\phi]~ \mathcal{O} ~e^{-S^{(\tau_0+\epsilon)}[\phi(\mathbf{x})]}, \\
&&\phi(\mathbf{x}\in \mathcal{D}_{\tau_0+\epsilon}), \\
&&\phi\left(\tau=\tau_0+\epsilon,x\right) = \varphi(x).
\end{eqnarray}
That is, now the path integral is over field configurations $\phi(\mathbf{x})$ for $\mathbf{x}\in \mathcal{D}_{\tau_0+\epsilon}$ and such that they coincide with $\varphi(x)$ at the boundary $\Sigma_{\tau_0+\epsilon}$ of $\mathcal{D}_{\tau_0+\epsilon}$, with
\begin{eqnarray}
S^{(\tau_0+\epsilon)}[\phi(\mathbf{x})] &\equiv&\int_{-\infty}^{\tau_0+\epsilon} d\tau \int dx~\mathcal{L}(\phi(\mathbf{x}), \partial_{\mu}\phi(\mathbf{x})),
\end{eqnarray}
Notice that the insertion $\mathcal{O}$ is the same as before. The new wavefunctional $\Psi'[\varphi(x)]$ corresponds to a vector $\ket{\Psi'}$ in the Hilbert space $\mathcal{H}(\Sigma_{\tau_0+\epsilon})$, but with the identification of Eq.\ \eqref{eq:identification} we can think of both states $\ket{\Psi}$ and $\ket{\Psi'}$ as belonging to the same Hilbert space $\mathcal{H}(\Sigma)$.

To compute the evolution from $\ket{\Psi}$ to $\ket{\Psi'}$ we will consider a change of coordinates as in Eq.\ \eqref{eq:Axx} where $\epsilon^{\mu} = \epsilon^{\mu}(\mathbf{x})$ is a smooth, infinitesimal function of $\mathbf{x} \in \mathcal{D}_{\tau_0}$ such that: (i) it is only non-zero in a small neighborhood of the boundary $\Sigma_{\tau_0}$ of $\mathcal{D}_{\tau_0}$ that does not contain 
the positions $\mathbf{x}_{1}, \cdots, \mathbf{x}_{n}$ of the insertions $\mathcal{O}$. (ii) when acting on time slice $\Sigma_{\tau_0}$, the coordinate change maps each point $(\tau_0,x)$ into the point $(\tau_0 + \epsilon, x)$ of the time slice $\Sigma_{\tau_0+\epsilon}$, that is
\begin{equation}
\epsilon^{\mu}(\tau_0,x) = \epsilon \xi^{\mu}(\tau_0,x) = \epsilon(1,0),
\end{equation}
where $\xi^{\mu}(\mathbf{x})$ had been introduced in Eq.\ \eqref{eq:xi1}.

We can now proceed to define a new field $\phi'(\mathbf{x})$ through $\phi'(\mathbf{x}')= \phi(\mathbf{x})$ as before, and study how the action functional $S[\phi(\mathbf{x})]$ changes under the coordinate and field transformations, as we did previously. This time, however, we must account for the change in the boundary conditions. Let $\phi(\mathbf{x})$ be a field configuration on $\mathcal{D}_{\tau_0}$ that fulfills the boundary conditions $\phi(\tau_{0},x) = \varphi(x)$ on $\Sigma_{\tau_0}$. Its action reads
\begin{eqnarray}
&&S^{(\tau_0)}[\phi(\mathbf{x})] \equiv \int_{-\infty}^{\tau_0} d\tau \int dx~\mathcal{L}(\phi(\mathbf{x}), \partial_{\mu}\phi(\mathbf{x}))  \label{eq:ASS1}\\
&=& \int^{\tau_0+\epsilon}_{-\infty} d\tau' \int dx' ~\left(1-\partial_{\mu}\epsilon^{\mu}(\mathbf{x}') \right) ~\times\label{eq:ASS2}\\
&&~~~ \mathcal{L}\left(\phi'(\mathbf{x}'), \partial'_{\mu} \phi'(\mathbf{x}') + (\partial'_{\mu}\epsilon^{\rho}(\mathbf{x}'))\partial'_{\rho}\phi'(\mathbf{x}')\right)~~~~\nonumber\\
&=& \int^{\tau_0+\epsilon}_{-\infty} d\tau \int dx ~\left(1-\partial_{\mu}\epsilon^{\mu}(\mathbf{x})\right)~\times \label{eq:ASS3}\\
&&~~~ \mathcal{L}\big(\phi'(\mathbf{x}), \partial_{\mu} \phi'(\mathbf{x}) + (\partial_{\mu}\epsilon^{\rho}(\mathbf{x})) \partial_{\rho} \phi'(\mathbf{x})\big)~~~~  \nonumber\\
 &=& \int^{\tau_0+\epsilon}_{-\infty} d\tau \int dx~\mathcal{L}\left(\phi'(\mathbf{x}), \partial_{\mu} \phi'(\mathbf{x})\right) + \label{eq:ASS4}\\
 &&~~ \int^{\tau_0}_{\infty} d\tau \int dx~\left(\partial_{\nu}\epsilon^{\mu}(\mathbf{x})\right)  \left( - \delta_{\mu}^{\nu} \mathcal{L} + \frac{\delta \mathcal{L}}{\delta \partial_\mu \phi'}\partial_{\nu} \phi'(\mathbf{x})\right) \nonumber \\
&=& S^{(\tau_0+\epsilon)}[\phi'(\mathbf{x})] + \int^{\tau_0}_{-\infty}\!\!\!\! d\tau \!\int \!dx\left(\partial_{\mu}\epsilon^{\nu}(\mathbf{x})\right) (T_E)^{\mu}_{\nu}(\mathbf{x}). \label{eq:ASS5}
\end{eqnarray}
In going from \eqref{eq:ASS1} to \eqref{eq:ASS2} we have explicitly accounted for the change in the integration domain. In the next step we simply relabeled the integration coordinates $\mathbf{x}$. In \eqref{eq:ASS3} we separated a first term which only depends on $\epsilon$ through the domain of integration of $\tau$, and a second term that collects all contributions linear in (the derivative of) $\epsilon^{\mu}(\mathcal{x})$ (notice that we have reset the integration limit of $\tau$ to $\tau_0$, but introducing corrections that are quadratic in $\epsilon$). Then in \eqref{eq:ASS5} we have identified the first term with the action of the field configuration $\phi'(\mathbf{x}) = \phi(\mathbf{x-\epsilon})$, which is defined on the domain $\mathbf{x} \in \mathcal{D}_{\tau_0+\epsilon}$ and fulfills the boundary conditions $\phi'(\tau_0+\epsilon,x) = \varphi(x)$, and the second term as an integral of the stress tensor.

Finally, we are ready to apply the same coordinate and field transformations to the path  integral expression of the wavefunctional $\Psi[\varphi(x)]$. We obtain:
\begin{eqnarray}
&&\Psi[\varphi(x)] \equiv \int [D\phi]~ \mathcal{O} ~e^{-S^{(\tau_0)}[\phi(\mathbf{x})]}\\
&=& \int [D\phi'] ~\mathcal{O} ~ e^{-S^{(\tau_0 + \epsilon)}[\phi'(\mathbf{x})]}\\ &\times& \left(1 - \int^{\tau_0}_{\infty}\!\!\!\! d\tau \!\int \!dx\left(\partial_{\mu}\epsilon^{\nu}(\mathbf{x})\right) (T_E)^{\mu}_{\nu}(\mathbf{x})\right).
\end{eqnarray}
Notice that above it is crucial that the insertions $\mathcal{O}$ are not affected by the change of coordinates. We also note that, since in this last expression the field $\phi'(\mathbf{x})$ is an integration variable, we can rename it $\phi(\mathbf{x})$. It then follows  that
\begin{eqnarray} \label{eq:AW4}
&&\Psi[\varphi(x)] = \int [D\phi] ~\mathcal{O} ~ e^{-S^{(\tau_0 + \epsilon)}[\phi(\mathbf{x})]} ~-\\ 
&& \int^{\tau_0}_{-\infty}\!\!\!\! d\tau \!\int \!dx \!\int [D\phi] ~\partial_{\mu}\left(\epsilon^{\nu} (\mathbf{x})(T_E)^{\mu}_{\nu}(\mathbf{x}) \right)\mathcal{O}  e^{-S^{(\tau_0)}[\phi(\mathbf{x})]} \nonumber\\
&&~~~~~~~~= \Psi'[\varphi(x)] -  \label{eq:AW5}\\
&& \int_{\Sigma_{\tau_0}}\!\!\!\! dx \sqrt{g_{11}}~ n_{\mu}(\mathbf{x})\epsilon^{\nu}(\mathbf{x}) \int [D\phi] (T_E)^{\mu}_{\nu}(\mathbf{x}) ~\mathcal{O} ~ e^{-S^{(\tau_0)}[\phi(\mathbf{x})]}. \nonumber 
\end{eqnarray}
In \eqref{eq:AW4} we have used the Ward identity in Eq.\ \eqref{eq:Ward} to extend the domain of the partial derivative to include the stress tensor, whereas in \eqref{eq:AW5} we have identified the first term in \eqref{eq:AW4} with the new wavefunctional and we have applied Stokes' theorem (namely $\int_{\mathcal{D}} d^2x~ \partial_{\mu} A^{\mu} = \int_{\partial \mathcal{D}} \epsilon_{\mu\nu}A^{\mu}dx^{\nu}$) to the second term in \eqref{eq:AW4}.
After introducing the field operator $\hat{\phi}(x)$ and its conjugate momentum $\hat{\pi}(x)$ on $\mathcal{H}(\Sigma_{\tau_0})$, and making the replacements $\phi(\tau_0,x) \rightarrow \hat{\phi}(x)$ and $\partial_{\tau} \phi(\tau_0,x) \rightarrow \hat{\pi}(x)$, Eqs.\ \eqref{eq:AW4}-\eqref{eq:AW5} can be seen to be equivalent to $\ket{\Psi} = \ket{\Psi'} + \epsilon Q\ket{\Psi}$, that is
\begin{eqnarray}
\ket{\Psi'} = \left(1 - \epsilon Q \right)\ket{\Psi},
\end{eqnarray}
where 
\begin{eqnarray}
Q &\equiv& - \frac{1}{\epsilon}\int_{\Sigma_{\tau_0}} dx~ \sqrt{g_{xx}(\mathbf{x})}~ n^{\mu}(\mathbf{x})~ \epsilon^{\nu}(\mathbf{x})~ (T_E)_{\mu\nu}(\mathbf{x}) ~~~\\
&=& - \int_{\Sigma_{\tau_0}} dx~ \sqrt{g_{xx}(\mathbf{x})}~ n^{\mu}(\mathbf{x})~\xi^{\nu}(\mathbf{x})~(T_E)_{\mu\nu}(\mathbf{x}). ~~~
\end{eqnarray}

\subsection{Example: free boson QFT in flat spacetime} 
\label{subsec:1I}

Here we briefly compare the lorentzian and euclidean QFT formalisms, with a free boson as a concrete example. We temporarily use subscripts ${}_L$ and ${}_E$ to differentiate between the two cases. 

Recall that in a 2d lorentzian QFT with bosonic field $\phi(\mathbf{x})$, where $\mathbf{x}=(t,x)$ are cartesian coordinates on Minkowski space-time with metric $\eta_{\mu\nu} = $diag$(-1,1) = \eta^{\mu\nu}$, the path integral $Z_{L}$ and the action functional $S_{L}$ read
\begin{eqnarray}
Z_L &=& \int [D\phi]~e^{i S_L[\phi(\mathbf{x})]}, \\
S_L[\phi(\mathbf{x})] &=& \int dtdx ~\mathcal{L}_L(\phi(\mathbf{x}), \partial_t\phi(\mathbf{x}), \partial_x\phi(\mathbf{x})),
\end{eqnarray}
where $\mathcal{L}_L$ is the lagrangian. Invariance of the action under space-time translations leads to the classically conserved stress tensor $(T_L)^{\mu}_{\nu}$,
\begin{equation}
(T_L)^{\mu}_{\nu} \equiv \frac{\delta \mathcal{L}_L}{\delta \partial_{\mu}\phi}\partial_{\nu} \phi - \delta^{\mu}_{\nu}\mathcal{L}_L,
\end{equation}
with energy and momentum densities $h_L$ and $p_L$, and classically conserved charges $H_L$ and $P_L$ given by
\begin{eqnarray}
h_L(\mathbf{x}) &\equiv& (T_L)^{0}_{0}(\mathbf{x}),~~~~~~H_L \equiv \int_{\Sigma_t} dx~ h_L(\mathbf{x}),~~~\\
p_L(\mathbf{x}) &\equiv& (T_L)^{0}_{1}(\mathbf{x}),~~~~~~P_L \equiv \int_{\Sigma_t} dx~ p_L(\mathbf{x}).
\end{eqnarray}
For a massive free boson we have
\begin{eqnarray}
\mathcal{L}_{L,b} &\equiv& \frac{1}{2}\left((\partial_t \phi)^2 - (\partial_x \phi)^2 - m^2 \phi^2 \right),\\
(T_{L,b})^{\mu}_{\nu} &=& -\eta^{\mu\alpha}\partial_{\alpha}\phi\partial_{\nu}\phi - \delta^{\mu}_{\nu}\mathcal{L}_{L,b},\\
h_{L,b} &=& \frac{1}{2}\left((\partial_t \phi)^2 + (\partial_x \phi)^2 + m^2\phi^2\right)\\
p_{L,b} &=& \partial_t\phi \partial_x \phi.
\end{eqnarray}

Let us now consider, as we did in most of this paper, a 2d euclidean QFT with 
with bosonic field $\phi(\mathbf{x})$, where now $\mathbf{x}=(\tau,x)$ are cartesian coordinates on the euclidean plane with metric $\delta_{\mu\nu} = $diag$(1,1) = \delta^{\mu\nu}$. The path integral $Z_{E}$ and the action functional $S_{E}$ read
\begin{eqnarray}
Z_E &=& \int [D\phi]~e^{-S_E[\phi(\mathbf{x})]}, \\
S_E[\phi(\mathbf{x})] &=& \int d\tau dx ~\mathcal{L}_E(\phi(\mathbf{x}), \partial_\tau\phi(\mathbf{x}), \partial_x\phi(\mathbf{x})),
\end{eqnarray}
where $\mathcal{L}_E$ is the lagrangian. Invariance under spacetime translations leads to the classically conserved stress tensor
\begin{equation}
(T_E)^{\mu}_{\nu} \equiv \frac{\delta \mathcal{L}_E}{\delta \partial_{\mu}\phi}\partial_{\nu} \phi - \delta^{\mu}_{\nu}\mathcal{L}_E,
\end{equation}
with energy and momentum densities $h_E$ and $p_E$, and classically conserved charges $H_E$ and $P_E$ given by
\begin{eqnarray}
h_E(\mathbf{x}) &\equiv& (T_E)^{0}_{0}(\mathbf{x}),~~~~~~H_E \equiv \int_{\Sigma_\tau} dx~ h_E(\mathbf{x}),~~~\\
p_E(\mathbf{x}) &\equiv& (T_E)^{0}_{1}(\mathbf{x}),~~~~~~P_E \equiv \int_{\Sigma_\tau} dx~ p_E(\mathbf{x}).
\end{eqnarray}
For a massive free boson we have
\begin{eqnarray}
\mathcal{L}_{E,b} &\equiv& \frac{1}{2}\left((\partial_\tau \phi)^2 + (\partial_x \phi)^2 + m^2 \phi^2 \right),\\
(T_{E,b})^{\mu}_{\nu} &=& \delta^{\mu\alpha}\partial_{\alpha}\phi\partial_{\nu}\phi - \delta^{\mu}_{\nu}\mathcal{L}_{E,b},\\
h_{E,b} &=& \frac{1}{2}\left((\partial_\tau \phi)^2 - (\partial_x \phi)^2 - m^2\phi^2\right),\\
p_{E,b} &=& \partial_\tau \phi \partial_x \phi.
\end{eqnarray}

Notice that the euclidean lagrangian $\mathcal{L}_E$ can be obtained from the lorentzian lagrangian $\mathcal{L}_L$ by replacing $\partial_t$ with $i\partial_{\tau}$ and adding a an overall minus sign, that is
\begin{equation}
\mathcal{L}_E(\phi(\mathbf{x}), \partial_{\tau} \phi(\mathbf{x}),\partial_{x} \phi(\mathbf{x})) = -\mathcal{L}_L(\phi(\mathbf{x}), i\partial_{\tau} \phi(\mathbf{x}),\partial_{x} \phi(\mathbf{x})). \nonumber
\end{equation}

In defining operators in the Hilbert space $\mathcal{H}(\Sigma_{t_0})$ or $\mathcal{H}(\Sigma_{\tau_0}) \cong \mathcal{H}(\Sigma_{t_0})$ at time $t=t_0$ or $\tau = \tau_0$, besides the field operator $\hat{\phi}(x)$ we also introduce its conjugate momentum $\hat{\pi}(x)$, which is the same in lorentzian and euclidean QFT and is defined as
\begin{equation}
\pi(x) \equiv \frac{\delta \mathcal{L}_L}{\delta \partial_t \phi} = i \frac{\delta \mathcal{L}_E}{\delta \partial_\tau \phi}.
\end{equation}
For instance, in the free boson QFT considered above, we have
\begin{equation}
\pi(x) \equiv \frac{\delta \mathcal{L}_{L,b}}{\delta \partial_t \phi} = i \frac{\delta \mathcal{L}_{E,b}}{\delta \partial_\tau \phi} = \partial_t \phi = i\partial_\tau \phi.
\end{equation}
Then, remaining with the free boson, the hamiltonian and momentum densities $\hat{h}_{L,b}$ and $\hat{p}_{L,b}$ and the hamiltonian and momentum operators $\hat{H}_{L,b}$ and $\hat{P}_{L,b}$, which are the generators of unitary transformations (in the Hilbert space $\mathcal{H}(\Sigma_{t_0}) \cong \mathcal{H}(\Sigma_{\tau_0})$) corresponding to time and space translations, are obtained from $h_{L,b}$, $p_{L,b}$, $H_{L,b}$ and $P_{L,b}$ by replacing $\phi(t_0,x) \rightarrow \hat{\phi}(x)$ and $\partial_{t}\phi(t_0,x) \rightarrow \hat{\pi}(x)$, and thus read
\begin{eqnarray}
\hat{h}_{L,b}(\mathbf{x}) &=& \frac{1}{2} \left(\hat{\pi}(x)^2 + (\partial_x \hat{\phi}(x))^2 + m^2\hat{\phi}(x)^2\right),~~~~~\\
\hat{p}_{L,b}(\mathbf{x}) &=&   \hat{\pi}(x) \partial_x \hat{\phi}(x),\\
\hat{H}_{L,b} &=& \int dx~\hat{h}_{L,b}(\mathbf{x}),~~~~~\\
\hat{P}_{L,b} &=& \int dx~ \hat{p}_{L,b}(\mathbf{x}).
\end{eqnarray}
The euclidean Hamiltonian and momentum densities the hamiltonian and momentum densities $\hat{h}_{E,b}$ and $\hat{p}_{E,b}$, and hamiltonian and momentum operators $\hat{H}_{E,b}$ and $\hat{P}_{E,b}$,which are the generators of euclidean time evolution and of space translations in the same Hilbert space, are obtained instead from $h_{E,b}$, $p_{E,b}$, $H_{E,b}$ and $P_{E,b}$ by replacing $\phi(\tau_0,x) \rightarrow \hat{\phi}(x)$ and $i\partial_{\tau}\phi(\tau_0,x) \rightarrow \hat{\pi}(x)$. Thus they read 
\begin{eqnarray}
\hat{h}_{E,b}(\mathbf{x}) &=& -\frac{1}{2} \left(\hat{\pi}(x)^2 + (\partial_x \hat{\phi}(x))^2 + m^2\hat{\phi}(x)^2\right),~~~~~\\
\hat{p}_{E,b}(\mathbf{x}) &=&  -i \hat{\pi}(x) \partial_x \hat{\phi}(x),\\
\hat{H}_{E,b} &=& \int dx~\hat{h}_{E,b}(\mathbf{x}),~~~~~\\
\hat{P}_{E,b} &=& \int dx~ \hat{p}_{E,b}(\mathbf{x}).
\end{eqnarray}

We conclude that lorentzian and euclidean densities/generators are related to each other as
\begin{eqnarray}
\hat{h}_{E,b} &=& -\hat{h}_{L,b},~~~~\hat{p}_{E,b} = - i\hat{p}_{L,b},\\
\hat{H}_{E,b} &=& -\hat{H}_{L,b},~~~\hat{P}_{E,b} = - i\hat{P}_{L,b}.
\end{eqnarray}
This is consistent with the Heisenberg picture operator evolutions
\begin{eqnarray}
\left[\hat{H}_{E,b}, \hat{\phi}\right] &=& -\left[\hat{H}_{L,b}, \hat{\phi}\right] = i\partial_t \hat{\phi} = -\partial_{\tau} \hat{\phi}, \\
\left[\hat{P}_{E,b}, \hat{\phi}\right] &=& -i\left[\hat{P}_{L,b}, \hat{\phi}\right] = -\partial_{x} \hat{\phi}. 
\end{eqnarray}
 
\section{Appendix: Refining a tensor network toward a strip of continuous path integral}
\label{sec:app_refinement}

\begin{figure*}
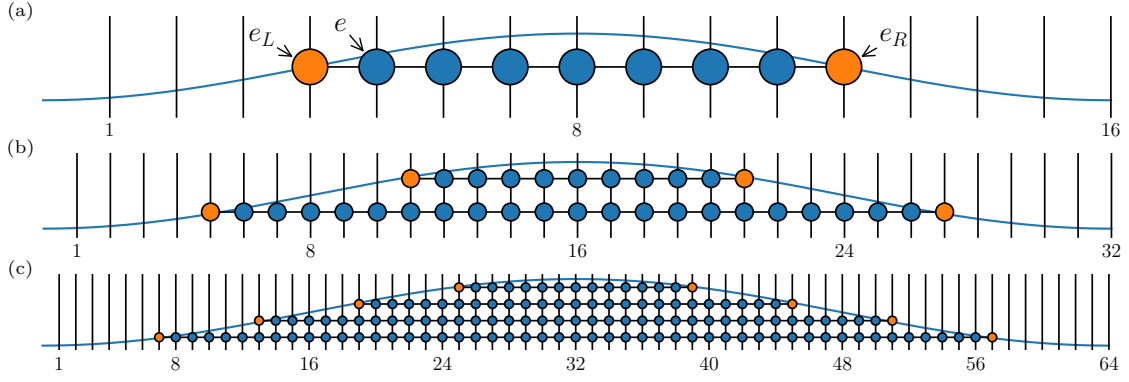

  \scriptsize
  (a)
  \raisebox{-\height}{\includegraphics[width=0.8\linewidth]{TN_eucl_cosine_diag_N16.pdf}}\\
  (b)
  \raisebox{-\height}{\includegraphics[width=0.8\linewidth]{TN_eucl_cosine_diag_N32.pdf}}\\
  (c)
  \raisebox{-\height}{\includegraphics[width=0.8\linewidth]{TN_eucl_cosine_diag_N64.pdf}}
\caption{\label{fig:TNs_cosine} Tensor networks, consisting of euclideons (blue) and euclideon smoothers (orange), implementing nonuniform euclidean time evolution. The sequence (a),(b),(c) represents refinement of a discretization of a corresponding strip of euclidean-time path integral on a flat cylinder $(\tilde \tau \in \mathbb{R}, \tilde x \in [0,2\pi))$, $\dee s^2 = \dee \tilde \tau^2 + \dee \tilde x^2$, bounded by timeslices $(0,x)$ and $(f(x),x)$, where $f(x) = \alpha(1 - \cos(x))$ for $\alpha=2\pi/32$.}
\end{figure*}

In this section we describe a sequence of tensor networks, illustrated in Fig.~\ref{fig:TNs_cosine}, whose actions as linear maps, on the low-energy states of a critical spin chain, approach the action of a strip of continuous euclidean-time path integral. For a CFT, the latter can be viewed as a conformal transformation whose matrix elements we can compute perturbatively in the thickness of the strip. We are therefore able to confirm that the tensor network maps are approaching the correct continuum limit by comparing their matrix elements with equivalent matrix elements of the conformal transformation.

Each tensor network is designed to implement nonuniform euclidean time evolution on a \emph{flat cylinder}, with the amount of euclidean time evolution at each point $x$ in space approximating, in the discretuum, the continuous function
\begin{equation} \label{eq:refex_cos_prof}
  f(x) \equiv \alpha(1 - \cos(m x)),
\end{equation}
where $\alpha > 0$ and $m\in\mathbb{Z}$. We carry out computations for the case $m=1$, for which the tensor networks are those illustrated in Fig.~\ref{fig:TNs_cosine} in the main text. Each tensor network has a different effective lattice spacing, with the sequence corresponding to a refinement of $f(x)$. As such we expect the action of the linear maps implemented by these tensor networks to approach that of a corresponding nonuniform strip of euclidean-time path integral. We now examine how to compute matrix elements of the latter.

\subsection{CFT matrix elements}
\label{sec:sup_ref_CFT}

Given a CFT Hilbert space $\mathcal{H}^{\CFT}(\Sigma_{\inn})$, $\mathcal{H}^{\CFT}(\Sigma_{\out})$ associated with timeslices $\Sigma_{\inn}$ and $\Sigma_{\out}$ of a CFT euclidean time path integral on a manifold $\mathcal{M}$, we wish to compute the action of the linear map $V$ given by the strip of path integral enclosed by $\Sigma_{\inn}$ and $\Sigma_{\out}$. We specialize to the particular case of a flat cylinder $\mathcal{M}=\mathbb{R} \times S^1$ with coordinates $\tilde \tau \in \mathbb{R}$, $\tilde x \in [0,2\pi)$ and metric
\begin{equation} \label{eq:refex_metric}
  \dee s^2 = \dee \tilde\tau^2 + \dee \tilde x^2
\end{equation}
and where the timeslices are defined by
\begin{align}
  \Sigma_{\inn}:& \quad \tilde\tau=0, \; \tilde x \in [0,2\pi) \\
  \Sigma_{\out}:& \quad \tilde\tau=f(\tilde x), \; \tilde x \in [0,2\pi),
\end{align}
where $f(\tilde x)$ is defined in Eq.~\eqref{eq:refex_cos_prof}.
Given an action for the path integral, this is enough to define the linear map $V$. Note, however, that the two timeslices have different proper lengths. In particular, differences in $\tilde x$ do not equal the corresponding proper distances on $\Sigma_{\out}$. If we compare states on the timeslices using the common $\tilde x$ coordinate, we are implicitly applying a local rescaling -- a Weyl transformation -- $W$ that matches up the internal geometry of one timeslice with that of the other. The effective map is thus $WV$. On the CFT Hilbert space, $W$ acts as a multiple of the identity and can be ignored, as discussed in the main text.

As discussed in the main text, the matrix elements $\langle \phi_\alpha | V | \phi_\beta\rangle$ for eigenstates $|\phi_\alpha \rangle$ of the CFT Hamiltonian $H_{\CFT}$ on $\Sigma_{\inn}$ are those of an appropriately chosen \emph{conformal transformation}
\begin{equation} \label{eq:refex_CThilb}
  V = e^{-Q},
\end{equation}
where the generator $Q$ can be expressed in terms of chiral and anti-chiral Virasoro generators
\begin{equation}
  Q = \sum_{n=-\infty}^{n=+\infty} \left[ a_n L_n + b_n \bar{L}_n \right].
\end{equation}

In the following, we fist show how to determine the generator of the conformal \emph{coordinate} transformation in terms of conformal generators obeying the Witt algebra. We then obtain the generator $Q$ of the corresponding transformation on the CFT Hilbert space by replacing elements of the Witt algebra with elements of the Virasoro algebra. Given irreducible representations of the Virasoro algebra belonging to a particular CFT, we apply the Virasoro commutator to compute, perturbatively, the desired matrix elements of $V$.

\subsubsection{Conformal coordinate transformation}
So that we may ultimately define, for the linear map $V$, a corresponding conformal transformation in terms of Virasoro generators, we need to work with a set of conformal coordinates that coincide with the coordinates $(\tilde\tau,\tilde x)$ on $\Sigma_{\inn}$. Fortunately, the $(\tilde \tau,\tilde x)$ coordinates are already conformal (indeed, the metric in these coordinates is flat). We can thus proceed to define a conformal coordinate transformation that, viewed as an active transformation, maps $\Sigma_{\inn}$ onto $\Sigma_{\out}$:
\begin{equation}
  (0,\tilde x) \rightarrow (f(\tilde x),\tilde x).
\end{equation}
A conformal coordinate transformation that achieves this is given by the holomorphic function
\begin{equation} \label{eq:refex_CTfunc}
  \omega'(\omega) \equiv f(-\ic \omega) + \omega,
\end{equation}
where $\omega \equiv \tilde \tau + \ic \tilde x$ and $\omega' \equiv \tilde \tau' + \ic \tilde x'$. In the new coordinates, the timeslice $\Sigma_{\out}$ is specified by
\begin{equation}
  \Sigma_{\out}: \quad \tilde \tau'=0, \; \tilde x' \in [0,2\pi).
\end{equation}

Let us briefly consider the metric in the $(\tilde \tau',\tilde x')$ coordinates: Given that $(\tilde \tau,\tilde x)$ are conformal coordinates, and that $(\tilde \tau',\tilde x')$ are related to $(\tilde\tau,\tilde x)$ by a conformal transformation (Eq.~\eqref{eq:refex_CTfunc}), the metric of Eq.~\eqref{eq:refex_metric} must take the form
\begin{equation}
  \dee s^2 = \Omega(\tilde \tau',\tilde x')^2 \left( \dee \tilde \tau'^2 + \dee \tilde x'^2 \right)
\end{equation}
for some function $\Omega(\tilde \tau',\tilde x')$, which we know to be nontrivial, since the proper distance along $\Sigma_{\out} \neq 1$. An example of a Weyl transformation that restores the feature that $\tilde x'$ measures the proper distance along $\Sigma_{\out}$ is
\begin{equation}
  W: \Omega(\tilde \tau',\tilde x') \rightarrow 1.
\end{equation}

Returning to the conformal coordinate transformation of Eq.~\eqref{eq:refex_CTfunc}, the next step is to find its generator. The generators of conformal coordinate transformations on the flat cylinder of unit radius are
\begin{align} \label{eq:Witt_gen}
  l_n = -e^{n\omega}\partial_\omega
\end{align}
and can be seen to realize the Witt algebra
\begin{align}
  [l_n,l_m] = (n-m)l_{n+m}.
\end{align}
Writing the transformation of Eq.~\eqref{eq:refex_CTfunc} as
\begin{equation} \label{eq:refex_CTgen_Witt}
  v \equiv e^{-q}
\end{equation}
so that $\omega' = v\omega$, we may determine the generator
\begin{equation}
  q = \sum_n a_n l_n
\end{equation}
perturbatively in the $\alpha$ parameter of Eq.~\eqref{eq:refex_cos_prof}. We expand the exponential in Eq.~\eqref{eq:refex_CTgen_Witt}, and seek $q$ so that $v\omega = \omega' + \mathcal{O}(\alpha^{k+1})$ for some chosen integer $k$. For $k=1$, this results in the generator coefficients
\begin{align}
  a_0 = \alpha, \quad a_m = a_{-m} = -\frac{\alpha}{2},
\end{align}
leading to the generator
\begin{align}
  q_{k=1} = \alpha\left(-1 + \cos(-\ic m\omega)\right) \partial_\omega,
\end{align}
so that
\begin{align}
  v_{k=1} = 1 + \alpha(1 - \cos(-\ic m\omega))\partial_\omega + \mathcal{O}(\alpha^2),
\end{align}
which matches Eq.~\eqref{eq:refex_CTfunc} up to $\mathcal{O}(\alpha^2)$ corrections.
For $k=4$, (shown in Fig.~\ref{fig:refinement}) the required coefficients are
\begin{align}
  a_0 &= \frac{37 \alpha^5 m^4}{1440}-\frac{\alpha^3 m^2}{12} + \alpha, \\
  a_m &= \frac{55 \alpha^5 m^4}{2304}-\frac{5\alpha^4 m^3}{96} -\frac{\alpha^3 m^2}{32}+\frac{\alpha^2  m}{4}-\frac{\alpha}{2}, \\
  a_{-m} &= \frac{55 \alpha^5 m^4}{2304}+\frac{5\alpha^4 m^3}{96}-\frac{\alpha^3 m^2}{32}-\frac{\alpha^2  m}{4}-\frac{\alpha}{2}, \\
  a_{2m} &= -\frac{13\alpha^5m^4}{288} -\frac{\alpha^4 m^3}{32}+\frac{\alpha^3 m^2}{8}-\frac{\alpha^2 m}{8},\\
  a_{-2m} &= -\frac{13\alpha^5m^4}{288} +\frac{\alpha^4 m^3}{32}+\frac{\alpha^3 m^2}{8}+\frac{\alpha^2 m}{8},\\
  a_{3m} &= -\frac{97\alpha^5m^4}{4608}+\frac{7\alpha^4m^3}{96}-\frac{5\alpha^3m^2}{96},\\
  a_{-3m} &= -\frac{97\alpha^5m^4}{4608}-\frac{7\alpha^4m^3}{96}-\frac{5\alpha^3m^2}{96},\\
  a_{4m} &= \frac{25\alpha^5 m^4}{576}-\frac{5\alpha^4 m^3}{192},\\
  a_{-4m} &= \frac{25\alpha^5 m^4}{576}+\frac{5\alpha^4 m^3}{192},\\
  a_{5m} &= a_{-5m} = -\frac{107\alpha^5 m^4}{7680},
\end{align}
where all $a_n$ not appearing above are set to zero.

Note that the generator of the conjugate coordinate transformation
\begin{equation} \label{eq:refex_CTfunc_conj}
  \bar\omega'(\bar\omega) = f(\ic \bar\omega) + \bar\omega
\end{equation}
is simply
\begin{equation}
  \bar q = \sum_n a_n^* \bar l_n.
\end{equation}

\subsubsection{Conformal transformation on the CFT Hilbert space}
We now proceed to \emph{quantize} the conformal coordinate transformation of Eq.~\eqref{eq:refex_CTfunc}. The quantized version of the Witt algebra is the Virasoro algebra (a central extension of the former). On a timeslice, such as $\Sigma_{\inn}$, of the flat Euclidean cylinder, two commuting copies of the Virasoro algebra are realized
\begin{eqnarray} \label{eq:Virasoro}
\big[L_n, L_m\big] &=& (n-m)L_{n+m} + \frac{c}{12}n(n^2-1)\delta_{n+m,0},~~~~~ \\
\big[ L_n, \bar{L}_m \big] &=& 0, ~~~\\
\big[\bar{L}_n, \bar{L}_m\big] &=& (n-m)\bar{L}_{n+m} + \frac{c}{12}n(n^2-1)\delta_{n+m,0},~~~~~
\end{eqnarray}
where $c$ is the \emph{central charge} and the chiral and anti-chiral Virasoro generators $L_n$ and $\bar{L}_n$ are the quantum equivalents of the Witt generators $l_n$ and $\bar{l}_n$. The quantization of a conformal transformation is therefore, roughly, the replacement of Witt generators with Virasoro generators in Eq.~\eqref{eq:refex_CTgen_Witt}. More precisely, we define the quantized transformation to be
\begin{equation} \label{eq:refex_CTgen_Vira}
  V = e^{-Q} = e^{-(Q_L + Q_R)}
\end{equation}
with the chiral and anti-chiral generators
\begin{align}
  Q_L &\equiv \sum_n a_n L_n - a_0\frac{c}{24} \label{eq:refex_Q}\\
  Q_R &\equiv \sum_n a_n^* \bar L_n - a_0^* \frac{c}{24}. \label{eq:refex_Qb}
\end{align}
The coefficients $a_n$ are those defined in the previous section for the coordinate transformation.

\subsubsection{Computation of matrix elements}

To compute matrix elements of $V$, we first expand the exponential in Eq.~\eqref{eq:refex_CTgen_Vira} to the same order $k$ used above to find the generator coefficients $a_n$.
Matrix elements of $V$ can then be approximately computed in terms of matrix elements of the various products of Virasoro generators in the expansion:
\begin{equation} 
  \langle \phi_1 | V | \phi_2 \rangle \approx \langle \phi_1 |\mathbb{I} - Q +\frac{1}{2}Q^2 + \dots + \frac{(-1)^k}{k!}Q^k | \phi_2\rangle.
\end{equation}

We now explain how to compute matrix elements of arbitrary products of Virasoro generators in terms of eigenstates of the CFT hamiltonian on the flat euclidean cylinder. The chiral and anti-chiral Virasoro algebras are realized on the Hilbert space of a timeslice $\tau=0$ of the cylinder as Fourier modes of the field operators $T(x)$ and $\bar{T}(x)$, the chiral and anti-chiral components of the stress tensor:
\begin{align}
  L_n &\equiv \frac{L}{(2\pi)^2} \int_{0}^{L} dx~e^{\ic\frac{2\pi}{L}nx } T(x) + \frac{c}{24}\delta_{n,0}, \label{eq:Ln}\\
  \bar{L}_n &\equiv \frac{L}{(2\pi)^2} \int_{0}^{L} dx~e^{-\ic\frac{2\pi}{L}nx } \bar{T}(x) + \frac{c}{24}\delta_{n,0} \label{eq:Ln_bar},
\end{align}
where $L$ is the circumference of the cylinder and in this case $L=2\pi$.
The stress tensor fields are related to the (lorentzian) energy and momentum densities as
\begin{eqnarray}
T(x) &\equiv&  2\pi\frac{h(x) + p(x)}{2},\\
\bar{T}(x) &\equiv& 2\pi\frac{h(x) - p(x)}{2},
\end{eqnarray}
where we recognize that the CFT hamiltonian and momentum operators can be expressed in terms of the Virasoro generators as
\begin{align}
H^{\CFT} &\equiv \int_0^{L} dx~h(x) = \frac{2\pi}{L} \left(L_0 +\bar{L}_0 - \frac{c}{12}\right), \label{eq:H_CFT}\\
P^{\CFT} &\equiv \int_0^{L} dx~p(x) =  \frac{2\pi}{L} \left(L_0 -\bar{L}_0\right).
\end{align}
As a result of this, given concrete representations of the chiral and anti-chiral Virasoro algebras corresponding to a particular CFT, we may compute matrix elements of products of Virasoro generators in terms of energy eigenstates.

An irreducible representation of the (chiral) Virasoro algebra is specified by the central charge $c$ and a conformal weight $h$. A particular CFT involves multiple such representations. Each has a state of highest weight $|h\rangle$, known as a primary state, from which all other states in the representation can be reached via the action of $L_n$ for $n<0$ (raising operators). A general (non-normalized) state can thus be written as a linear combination of states of the form
\begin{equation}\label{eq:refex_CFTstate}
  |\phi^{\mathrm{chiral}}\rangle = L_{n_1} L_{n_2} \dots L_{n_q} |h\rangle,
\end{equation}
with $q\in \mathbb{Z}$ and $n_j < 0$.
The primary state is annihilated by all $L_n$ with $n>0$ and satisfies
\begin{equation}
  L_0 |h\rangle = h |h\rangle.
\end{equation}
By Eq.~\eqref{eq:Virasoro} it follows that the states of Eq.~\eqref{eq:refex_CFTstate} are also eigenstates of $L_0$.
Together with the Virasoro commutator of Eq.~\eqref{eq:Virasoro}, this is enough to compute the matrix element
\begin{equation}
  \langle \phi^{\mathrm{chiral}}_1 | X | \phi^{\mathrm{chiral}}_2 \rangle
\end{equation}
of any product of Virasoro generators $X$. For more details see~\cite{Conformal}. This includes the norm of each state of the form $|\phi^{\mathrm{chiral}}\rangle$.

As is clear from Eq.~\eqref{eq:H_CFT}, the eigenstates of $H^{\CFT}$ are simultaneous eigenstates of $L_0$ and $\bar{L}_0$ and are (linear combinations of) states descended, by the action of chiral and anti-chiral Virasoro generators, from a primary state $|h,\bar h\rangle$ with chiral and anti-chiral conformal weights $h$ and~$\bar h$:
\begin{equation}
  |\phi^{\CFT}\rangle = L_{n_1} \dots L_{n_q} \bar{L}_{\bar{n}_1} \dots \bar{L}_{\bar{n}_{\bar{q}}} |h,\bar h\rangle,
\end{equation}
Since the two Virasoro algebras commute, computing matrix elements of such states is a straightforward extension of computing them for the states $|\phi^{\mathrm{chiral}}\rangle$ of Eq.~\eqref{eq:refex_CFTstate}.

\subsection{Tensor Network matrix elements}

We have a sequence of tensor networks, each implementing a linear map, which we expect to approximate the action of the path integral strip discussed in section~\ref{sec:sup_ref_CFT}. Each tensor network implements a linear map between spin chain Hilbert spaces associated with lattices $\mathcal{L}_{\inn}$ and $\mathcal{L}_{\out}$. Just as with the path integral, the tensor networks can be thought of as having been ``cut'' from a tensor network for the euclidean path integral on the full flat cylinder. The linear map associated with moving in the $\tilde \tau$ direction is the transfer matrix $\mathcal{T}$, which consists of a complete periodic row of euclideons. Energy eigenstates can be obtained by diagonalizing $\mathcal{T}$ or, equivalently, a generating spin chain hamiltonian $H \sim -\log(\mathcal{T})$. For a critical system with emergent lorentz invariance, each low-energy eigenstate $|\phi_j\rangle$ of $H$ can be related to an eigenstate $|\phi^{\CFT}_j\rangle$ of a corresponding CFT hamiltonian $H^{\CFT}$ for the CFT describing the universality class of the spin chain:
\begin{equation} \label{eq:refex_state_equiv}
  |\phi_j\rangle \sim |\phi^{\CFT}_j\rangle
\end{equation}
Numerical procedures for doing this, using lattice analogues of the Virasoro generators, are discussed in \cite{LatVira, puMPS, Conformal}. To compare the action of the tensor network maps $\mathrm{TN}$ with the path-integral linear map $V$, we compute matrix elements of each in terms of the eigenstates of the spin chain hamiltonian or the corresponding CFT hamiltonian, respectively. We restrict the comparison to a low-energy subspace, within which the effects of the lattice UV cutoff are sufficiently benign.

Computing matrix elements of a tensor network in terms of the eigenstates of a spin chain hamiltonian is conceptually simple. We first diagonalize the hamiltonian $H$, selecting a set of low-energy eigenstates $|\phi_j\rangle$. We may then use the techniques of \cite{LatVira, puMPS} to associate each energy eigenstates with corresponding eigenstates of $H^{\CFT}$. Finally, we compute the overlaps
\begin{equation} \label{eq:refex_TNels}
  \mathrm{TN}_{jk} \equiv \langle \phi_j | \mathrm{TN} | \phi_k \rangle,
\end{equation}
to be compared with the path integral matrix elements
\begin{equation}
  V_{jk} \equiv \langle \phi_j^{\CFT} | V | \phi_k^{\CFT} \rangle.
\end{equation}

\subsubsection{Periodic Matrix Product States}
In this case, we must compute matrix elements $\mathrm{TN}_{jk}$ for lattices with $N=16,32,64$ sites, which takes us into the regime where the full spin-chain Hilbert space $\mathcal{H}(\Sigma_{\inn}) \cong (\mathbb{C}^2)^{\otimes N}$ is so large that general states can no longer be stored in computer memory. Instead, we approximately represent states using a class of tensor network states known as periodic Matrix Product States \cite{MPS, pMPS} (pMPS). These states have the form
\begin{equation}
  |\psi(A)\rangle \equiv \sum_{s_1 \dots s_N} \tr\left(A_1^{s_1} A_2^{s_2} \dots A_N^{s_N}\right) |s_1 s_2 \dots s_N\rangle,
\end{equation}
where $|s_1 s_2 \dots s_N\rangle$ is an element in a chosen basis for the spin chain, and are parameterized by $N$ tensors $A_j$, each with dimension $d \times D \times D$, such that $A_j^s$ is a $D \times D$ matrix and $d$ is the dimension of the Hilbert space of one lattice site (in this case $d=2$). We represent a pMPS using tensor network notation as
\begin{equation}
  |\psi\rangle = \includegraphics[width=0.8\linewidth]{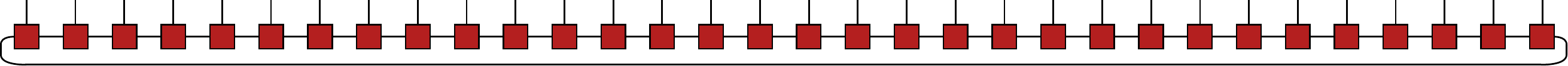},
\end{equation}
where in this example $N=32$.

The \emph{bond dimension} $D$ of a pMPS is a parameter that controls the amount of entanglement that can be represented \cite{MPS}. It is known that MPS can efficiently approximate ground states of critical systems \cite{MPS_crit}, in that the bond dimension required to maintain a given accuracy grows polynomially in $N$. In \cite{puMPS}, evidence is provided showing that excited states of critical systems can also be approximated efficiently.

\subsubsection{pMPS algorithms}

We find approximate ground states variationally by minimizing the energy within a translation-invariant class of pMPS, called periodic uniform MPS (puMPS)~\cite{pMPS}. We then approximate excited states by diagonalizing the effective hamiltonian in the tangent space, at the ground state, of the variational manifold of puMPS states~\cite{pMPS}. The techniques of~\cite{LatVira}, used to establish Eq.~\eqref{eq:refex_state_equiv}, can be applied efficiently to such states, as shown in~\cite{puMPS}. The algorithms used have a computational cost that scales, at worst, as $\mathcal{O}(N D^6)$.

\begin{widetext}
For the matrix elements $\mathrm{TN}_{jk}$, we also require a method to efficiently compute the inner product of Eq.~\eqref{eq:refex_TNels}.
The tensor networks of Fig.~\ref{fig:TNs_cosine} can be decomposed into horizontal rows of euclideons (terminated by smoothers) and applied row by row to a pMPS by contracting each euclideon (or smoother) with the adjacent MPS tensor, combining the horizontal indices of the pMPS with the corresponding horizontal indices of the euclideons (dimension $\chi$), to form a new MPS tensor with dimension $d \times D\chi \times D\chi$. Note that this is precisely the application of a Matrix Product Operator (MPO) to the state \cite{MPS}. Under this procedure, the maximum bond dimension reached is $\chi^r D$, where $r$ is the depth of the euclideon tensor network (the number of rows). The application of the tensor network of Fig.~\ref{fig:TNs_cosine}(b) to a pMPS of $N=32$ sites can be illustrated as
\begin{align}
  \mathrm{TN}|\psi\rangle &= \includegraphics[width=0.7\linewidth]{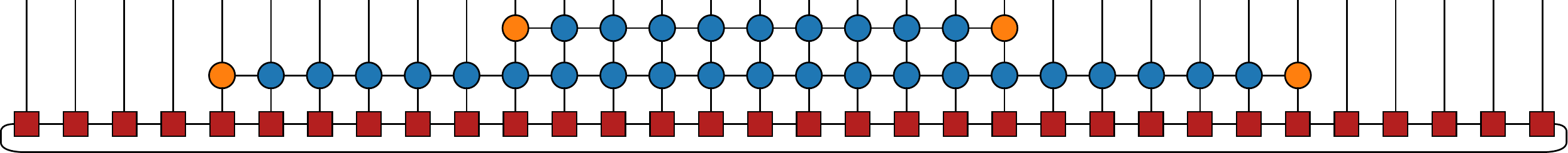} \label{eq:refex_TNpMPS1}\\
                          &= \includegraphics[width=0.7\linewidth]{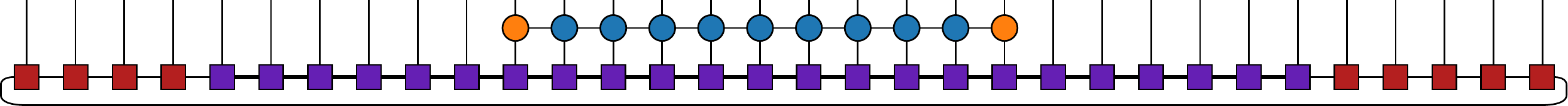} \label{eq:refex_TNpMPS2}\\
  &= \includegraphics[width=0.7\linewidth]{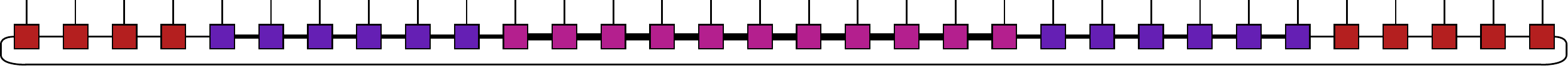} \label{eq:refex_TNpMPS3}
\end{align}
In Eqs.~\eqref{eq:refex_TNpMPS2} and~\eqref{eq:refex_TNpMPS3}, we indicate the larger effective bond dimensions of the intermediate and final pMPS by thicker lines and the inhomogeneity of the tensors by different colors.
\end{widetext}

Although this method is viable in principle, for $D$ already near the computationally tolerable limit the practicability of the inner product computation depends crucially on how large $\chi^r$ becomes for a given $\mathrm{TN}$. This is because the cost of computing an inner product of two pMPSs of bond dimension $D$ and $D'$ scales as $\mathcal{O}(N D^2 D'^3)$, where $D' > D$ is the larger of the two bond dimensions. Hence in our case we have $\mathcal{O}(N D^5 \chi^{3r})$. For the tensor networks of Fig.~\ref{fig:TNs_cosine} the maximum depth is $r=4$, so that even for $\chi = 2$ computing the inner product exactly is three orders of magnitude slower than the case $D=D'$. To mitigate this cost, we perform a \emph{compression} step after each application of a row of euclideons (MPO) to the pMPS.

To compress a pMPS parameterized by tensors $A_j'$, with maximum bond dimension $\chi D$, we seek tensors $\tilde A_j$ parameterizing a pMPS of bond dimension $\tilde D < \chi D$ that minimize the difference between the two states
\begin{equation}
  \left| |\psi(\tilde A)\rangle - |\psi(A)\rangle \right|^2.
\end{equation}
This can be done at cost $\mathcal{O}(N^2 \tilde D^2 D^3 \chi^3)$ \cite{pMPS}. With $\tilde D = D$, this is an improvement on the procedure without compression as long as $\chi^{3r-3}>N$, which is the case for the tensor network of Fig.~\ref{fig:TNs_cosine}(c), where $N=64$ and $r=4$. We observe that, even with $\tilde D - D$ the compression procedure achieves good fidelity. This makes sense, since the bulk of each row of euclideons is just a piece of the transfer matrix $\mathcal{T}$ and eigenstates of $H$ are also eigenstates of $\mathcal{T}$.

We also use compression on our excited states, which we obtain as puMPS tangent vectors \cite{pMPS}. The tangent vectors are given by a sum over $N$ pMPSs, each with bond dimension $D$, but can also be represented as a single pMPS with bond dimension $2D$. To avoid an extra factor $N^2$ in computations, we use the latter form, applying compression to reduce the bond dimension back to $D$, which we observe can be achieved to very good accuracy for low-energy eigenstates of the critical spin chain we used for testing. 

\subsection{Computations performed for the numerical comparison in the main text}

Having described how to compute matrix elements of the path-integral linear map $V$ and of the tensor networks $\mathrm{TN}$ of Fig.~\ref{fig:TNs_cosine}, as well as how to compare the two based on an identification of spin chain energy eigenstates with CFT energy eigenstates on the input timeslice $\Sigma_{\inn}$, we may perform these calculations for a particular critical spin chain.

We choose, for simplicity, the critical transverse-field Ising model, which has
\begin{equation}
  H = -\sum_{j=1}^N \left[\sigma^Z_j \sigma^Z_{j+1} + \sigma^X\right],
\end{equation}
where the Hilbert space of one lattice site is $\mathbb{C}^2$ and $\sigma^X$ and $\sigma^Z$ are Pauli matrices. The euclideon tensors $e$ for the Ising model can be derived from the Boltzmann weights for a plaquette of the 2D classical Ising model partition function
\begin{equation}
  e(s_1,s_2,s_3,s_4) = \exp((s_1s_2 + s_2s_3 + s_3s_4 + s_4s_1)\beta_c)
\end{equation}
where $s_j\in [-1,1]$ and $\beta_c = \log(1+\sqrt{2})/2$ is the critical inverse temperature. The euclideon tensor with basis compatible with $H$, in terms of tensor indices running from 1 to 2, is $e_{jklm} = e((-1)^j,(-1)^k,(-1)^l,(-1)^m)$.
The euclideon smoothers, needed to terminate a row of euclideons in tensor networks $\mathrm{TN}$, are derived using the methods of \cite{Conformal}.

For the tensor network matrix elements, we first compute approximate eigenstates of $H$ for system sizes $N=16,32,64$ corresponding to the tensor networks. For $N=16$ we use exact diagonalization of $H$ with matrix-free methods. For $N=32,64$ we use the puMPS techniques described in \cite{puMPS} with bond dimensions $D=18,26$. To compute matrix elements of the tensor networks, we compress tangent vectors to single pMPSs with bond dimension $D$ and employ compression with $\tilde D = D$ after applying each row of euclideons to a pMPS, as described above.

\begin{figure}
  \includegraphics[width=\linewidth]{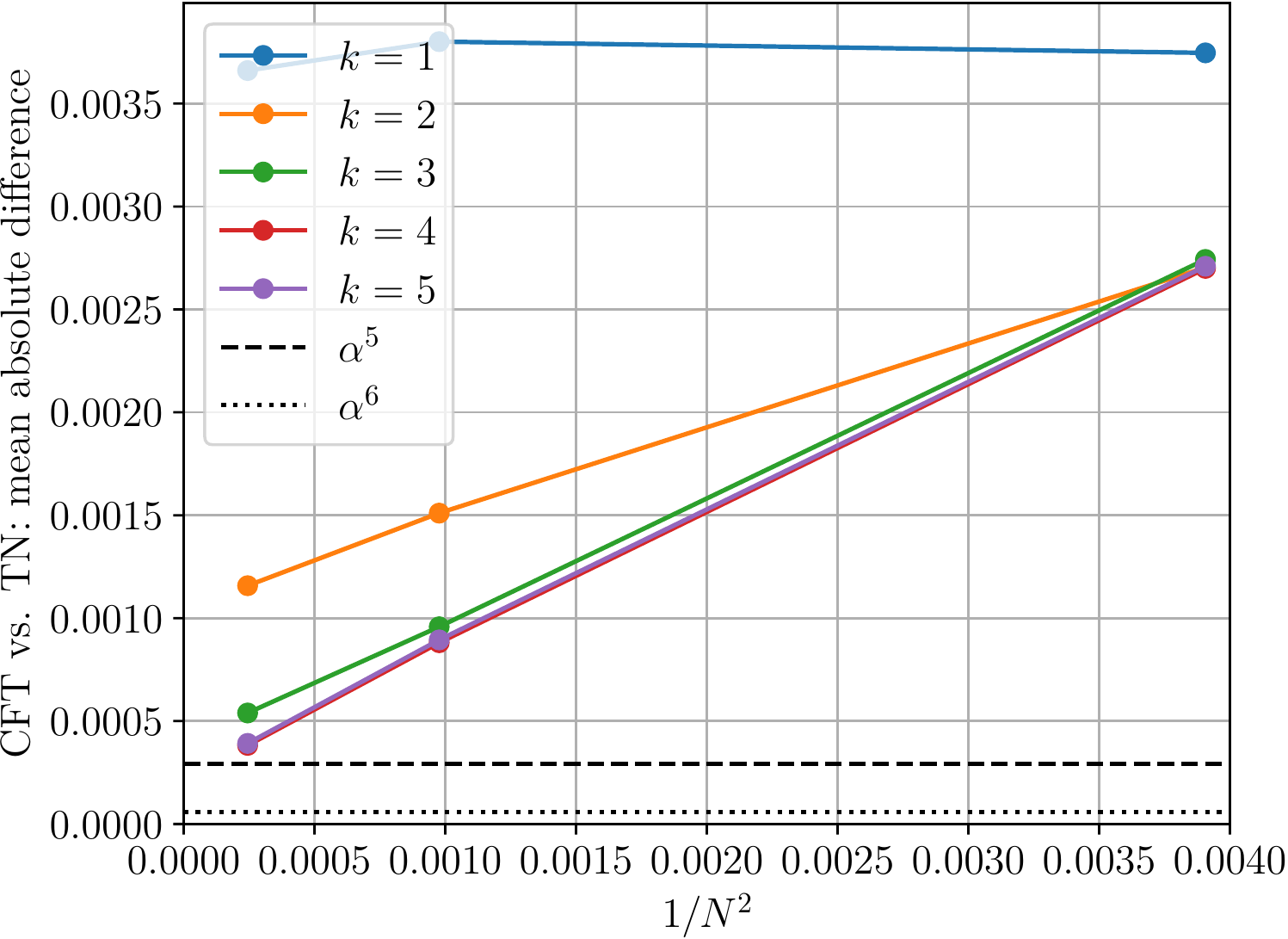}
  \caption{\label{fig:ref_err_app} Mean absolute difference (Eq.~\eqref{eq:mean_err}) between matrix elements of the tensor network maps of Fig.~\ref{fig:TNs_cosine} and the corresponding euclidean path integral map $V$, the matrix elements of which are computed up to order $k$. We also plot the values $\alpha^5$ and $\alpha^6$ as an indicator of the error we expect from the truncation to orders $k=4$ and $k=5$, respectively. Note that at this order of magnitude errors from the use of pMPS to approximate states also become relevant, which may help explain why the curves for $k=4$ and $k=5$ look very similar.}
\end{figure}

The emergent low-energy physics of the critical Ising model is described by the Ising CFT \cite{CFT}, which has central charge $c=1/2$. In the absence of defects, the following irreps of the Virasoro generators are present in Eqs.~\eqref{eq:Ln} and~\eqref{eq:Ln_bar}:
\begin{align}
  \mathbb{I}:& \quad h=\bar h=0 \\
  \sigma:& \quad h=\bar h=1/16 \\
  \varepsilon:& \quad h=\bar h=0.5
\end{align}
This is enough information to compute the matrix elements of the path integral map $V=e^{-Q}$, where we derive the generator $Q$ up to order $k$ in $\alpha$. For the tensor networks of Fig.~\ref{fig:TNs_cosine}, and a cylinder with circumference $2\pi$, we have $\alpha=2\pi/32$ and $m=1$ in Eq.~\eqref{eq:refex_cos_prof}. We carry out these computations for $k\le 5$.

The results of the comparison are shown for multiple values of $k$ in Fig.~\ref{fig:ref_err_app} in terms of the mean absolute difference between matrix elements
\begin{equation} \label{eq:mean_err}
  \frac{1}{M^2} \sum_{j=1}^{M}\sum_{k=1}^{M} \left| \mathrm{TN}^{(N)}_{jk} - V_{jk} \right|,
\end{equation}
where $M$ is the number of energy eigenstates for which we compute the matrix elements. In this case, $M=41$. We observe that the mean error tends toward a value compatible with zero, given an approximation of the error made in the perturbative expansion used to compute the generator $Q$ of $V$.

In Fig.~\ref{fig:refinement} of the main text, we plot the mean absolute difference within each conformal tower, and of the tower-mixing matrix elements, separately. It is interesting to note that the error for the identity tower does not converge to zero as well as the other towers. This may be an artifact of there being far fewer identity-tower matrix elements in the low-energy subspace consisting of the first $41$ eigenstates. In particular, a greater proportion of these matrix elements involve higher-energy states that are less well-approximated by the puMPS tangent vectors we use to approximate excited states.


\begin{thebibliography}{99}
\bibitem{rule1} A finite conformal map $V=e^{-Q}$ may in general have alternative descriptions of the form $V = e^{-Q'}$ in terms of a different generator $Q'\not = Q$ (or even as a path-ordered exponential $V = \mathcal{P}\exp(-\int_0^1 d\tau ~Q'(\tau))$ for a time dependent generator $Q'(\tau)$), each leading to an alternative geometry. This is expected: in a discrete network we cannot resolve the geometry at distances shorter than the spacing between nearest neighbour tensors. However, by analysing a very long strip, the linear map becomes infinitesimal, $e^{-Q} \approx \mathbb{I}-Q$, and we can then resolve these ambiguities by identifying the unique generator $Q$. For instance, a periodic layer $\mathcal{T}$ made of $N$ euclideons implements the map $e^{-H}$, where $H = \frac{2\pi}{N}\left(L_0 + \bar{L}_0 - c/12\right)$. Thus, for large $N$ its action on low energy states is $e^{-2\pi/N(\Delta_\alpha - c/12)} \approx 1 - 2\pi/N(\Delta_\alpha - c/12)$, where $\Delta_{\alpha}$ is the scaling dimension of the state. We can then unambiguously assign the generator $H$ (and thus the metric in Eq.\ \eqref{eq:gTN1}) to $\mathcal{T}$.
\bibitem{Weyl} In a CFT, the linear map $V$ implemented by the path integral is sensitive to a Weyl rescaling through a multiplicative constant given by the exponential of the Liouville action \cite{PathIntegral1,PathIntegral2}. This multiplicative constant, which does not affect correlation functions (since the same factor appears in the partition function with and without field insertions), is not reproduced by the tensor network \cite{TNRlocal} and will be ignored in our current discussion.
\end{thebibliography}
\end{document}